\documentclass[12pt,superscriptaddress,amsmath,amssymb,nofootinbib,longbibliography,preprint]{revtex4-1}

\usepackage{graphicx}
\usepackage{dcolumn} 
\usepackage{bm}
\usepackage{amssymb}
\usepackage{amsmath}
\usepackage{epsfig}    
\usepackage{color}
\usepackage{slashed}
\usepackage{hyperref}
\allowdisplaybreaks[1]
\begin{document} 
\date{\today}
\title{Flavor Dependent $U(1)$ Symmetric Zee Model\\ with a Vector-like Lepton}
\preprint{OU-HET-1080, CTP-SCU/2021004}

\author{Toshinori Matsui}
\email{matsui@kaist.ac.kr}
\affiliation{Department of Physics, Korea Advanced Institute of Science and Technology, 291 Daehak-ro, Yuseong-gu, Daejeon 34141, Republic of Korea}

\author{Takaaki Nomura}
\email{nomura@scu.edu.cn}
\affiliation{College of Physics, Sichuan University, Chengdu 610065, China}

\author{Kei Yagyu}
\email{yagyu@het.phys.sci.osaka-u.ac.jp}
\affiliation{Department of Physics, Osaka University, Toyonaka, Osaka 560-0043, Japan}

\begin{abstract}

We extend the Zee model by introducing a vector-like lepton doublet and a flavor dependent global $U(1)$ symmetry. 
Flavor changing neutral currents in the quark sector can be naturally forbidden at tree level due to the $U(1)$ symmetry, 
while sufficient amount of lepton flavor violation is provided to explain current neutrino oscillation data. 
In our model, additional sources of CP-violation appear in the lepton sector, but their 
contribution to electric dipole moments is much smaller than current experimental bounds due to the Yukawa structure constrained  by the $U(1)$ symmetry. 
We find that there is a parameter region where the strongly first order electroweak phase transition can be realized, 
which is necessary for the successful scenario of the electroweak baryogenesis in addition to new CP-violating phases. 
In the benchmark points satisfying neutrino data, lepton flavor violation data and the strongly first order phase transition, 
we show that an additional CP-even Higgs boson $H$ mainly decays into a lighter CP-odd Higgs boson $A$, i.e., $H \to AZ$ or $H \to AA$
with a characteristic pattern of lepton flavor violating decays of $A$.

\end{abstract}

\maketitle

\section{Introduction}

Neutrino oscillations are the phenomena which indicate clear evidence for the necessity of physics beyond the standard model (SM). 
From various measurements, it has been known that masses of neutrinos have to be much smaller than those of charged fermions, e.g., ${\cal O}(10^{12})$ times smaller than the top quark mass. 
This strongly suggests that the neutrino mass is generated by a different mechanism from that for charged fermions, i.e., Dirac masses via Yukawa interactions with a Higgs doublet field. 
In such context, Majorana masses for neutrinos can be a good candidate, which are effectively described by the dimension five Weinberg operator~\cite{Weinberg:1979sa}.  
The question is then how the Weinberg operator can be written in terms of the ultra-violet physics. 
The simplest example has been known as the type-I seesaw mechanism~\cite{Minkowski:1977sc,Yanagida:1980xy,Mohapatra:1979ia}, 
where only right-handed neutrinos are sufficient to be added to the SM. 
This, however, requires huge Majorana masses of the right-handed neutrinos, typically 
order ${\cal O}(10^{14})$ GeV assuming order one Dirac Yukawa couplings, so that direct detections for such heavy particles are quite challenging. 

As an alternative direction, 
we can consider a scenario where neutrino masses are generated via quantum effects by which new particles are not needed to be super heavy. 
This idea has originally been realized in the model proposed by A.~Zee~\cite{Zee:1980ai} (Zee model), 
where neutrino masses are generated at one-loop level\footnote{For the other models with radiatively induced neutrino masses, see the recent review paper~\cite{Cai:2017jrq}. }. 
In the Zee model\footnote{We call the model with a $Z_2$ symmetry as the Zee model which is sometimes called as the Zee-Wolfenstein model~\cite{Wolfenstein:1980sy}. }, 
right-handed neutrinos are not introduced, while the Higgs sector is extended by adding 
an additional Higgs doublet and charged singlet fields, by which the lepton number is explicitly broken via scalar interactions. 
The Zee model predicts a characteristic structure of the mass matrix for Majorana neutrinos, e.g., all the diagonal elements to be zero. 
Such a strong prediction ironically turns out to kill the model itself, because of the contradiction with the observed neutrino oscillations~\cite{Koide:2001xy,Frampton:2001eu,He:2003ih}. 
Therefore, modifications or extensions of the Zee model are inevitable. 

The simplest modification would be just not imposing a $Z_2$ symmetry which is originally introduced to avoid tree level 
flavor changing neutral currents (FCNCs) mediated by neutral Higgs bosons. 
In this case, both Higgs doublet fields can couple to each type of fermions, i.e., up-type quarks, down-type quarks and charged leptons, so that 
we obtain sufficient sources of lepton flavor violations in order to explain current neutrino data~\cite{He:2003ih,AristizabalSierra:2006ri,He:2011hs,Herrero-Garcia:2017xdu}. 
However, this requires unnatural fine-tunings in the quark sector to avoid various constraints from flavor changing processes such as the $B$-$\bar{B}$ mixing. 
Recently, in Ref.~\cite{Nomura:2019dhw} another possibility has been proposed, where a global $U(1)'$ symmetry is introduced instead of the $Z_2$ symmetry\footnote{The Zee model has also been extended by introducing 
a supersymmetry~\cite{Kanemura:2015maa}, an $A_4$ symmetry~\cite{Fukuyama:2010ff} and an $SU(3)_c \times SU(3)_L \times U(1)_X$ symmetry~\cite{Das:2020pai}. }. 
Taking flavor dependent charge assignments for lepton fields, 
we can obtain additional sources of lepton flavor violations, and at the same time matrices for quark Yukawa interactions are diagonal. 

In this paper, 
we clarify that the minimal Zee model with the $U(1)'$ symmetry cannot explain current neutrino oscillation data\footnote{We find an error in the structure of the neutrino mass matrix in Ref.~\cite{Nomura:2019dhw}.  }. 
We thus add a vector-like lepton doublet, as one of the simplest extensions, to the model, in which 
we assume a weak mixing between the new vector-like field and the SM leptons to avoid large contributions to charged lepton flavor violating (CLFV) decays. 
In this extension, an anti-symmetric Yukawa matrix among the lepton doublets and the charged singlet scalar becomes a $4\times 4$ form including six independent complex parameters which 
can be analytically solved in terms of the neutrino parameters driven by experiments.
This extended model also provides 
new sources of CP-violation in the lepton sector, which is well motivated for the explanation of baryon asymmetry of Universe~\cite{Sakharov:1967dj}. 
Interestingly, it is clarified that effects of the CP-violating phases on the electron electric dipole moment (EDM)
are negligibly small because of the structure of the Yukawa matrices constrained by the $U(1)'$ symmetry. 
We then study the electroweak phase transition, and find a region of the parameter space where 
the strongly first order phase transition (FOPT) is realized which is needed for the successful scenario of the electroweak baryogenesis~\cite{Kuzmin:1985mm,Shaposhnikov:1987tw}.
In addition, we discuss collider signatures of our model, particularly focusing on Higgs boson decays into a flavor violating lepton pair
in the  benchmark parameter points which satisfy neutrino data, CLFV data and the strongly FOPT. 

This paper is organized as follows. 
In Sec.~\ref{sec:model}, we define our model, and give the Yukawa interaction terms and the Higgs potential. 
Constraints from perturbative unitarity and vacuum stability are also discussed. 
In Sec.~\ref{sec:neutrino}, we discuss neutrino masses which are generated at one-loop level, 
and numerically show the necessity of the extension of the minimal Zee model in order to 
reproduce the current neutrino oscillation data.
Sec.~\ref{sec:flavor} is devoted for the discussion of various constraints from flavor experiments such as the electron EDM and CLFV decays. 
Collider phenomenologies are then discussed in Sec.~\ref{sec:collider}, particularly focusing on production and decay of additional neutral Higgs bosons at the LHC. 
In Sec.~\ref{sec:ewpt}, we show the electroweak phase transition as a cosmological consequences in our model. 
Conclusions are given in Sec.~\ref{sec:conclusion}. 
In Appendix, we give the approximate formulae for each element of the anti-symmetric matrix $F$ (Appendix~\ref{sec:fmat}), 
explicit expressions for the amplitude of the CLFV decays (Appendix~\ref{sec:LFV_amp}) and 
those for the effective potential at finite temperature (Appendix~\ref{sec:finiteT}). 

\section{Model\label{sec:model}}

We briefly review the Zee model with a  global $U(1)'$ symmetry which has been proposed in Ref.~\cite{Nomura:2019dhw}. 
The content of the scalar sector is the same as the original Zee model, which is composed of 
two isospin doublet fields $\Phi_1$ and $\Phi_2$ and a pair of singly-charged scalar singlets $S^\pm$. 
This model, however, cannot explain current neutrino oscillation data, as it will be shown in the next section. 
One of the simplest extensions is the introduction of a vector-like lepton doublet $L^4 \equiv (\nu_{T}, T^-)^T$ to the model. 
In this section, we first discuss Yukawa interactions with $L^4$, and then consider the Higgs potential.  

\subsection{Yukawa Interactions}

The most general form of the Lagrangian for the lepton sector is given by 
\begin{align}
\mathcal{L}_{\rm lep} & = - M'\bar{L}^4L^4
-\left[ \bar{L}_L^A (\tilde{Y}_1)^{AB} \Phi_1 \ell_R^B + \bar{L}_L^A(\tilde{Y}_2)^{AB} \Phi_2 \ell_R^B 
+\overline{L_L^{cA}}(\tilde{F})^{AB}(i\tau_2)L_L^BS^+ + \text{h.c.}\right], \label{eq:yukawa}
\end{align}
where $L_L$ $(\ell_R)$ are the left-handed (right-handed) lepton fields. 
The indices $A$ and $B$ ($=1,\dots,4$) represent the flavor with $L_L^{1\text{--}3}$ and $\ell_R^{1\text{--}3}$ 
to be identified with the SM lepton fields, and  $\ell_R^4 \equiv T_R$ which is the charged component of $L_R^4$.
The supscript $c$ denotes the charge conjugation. 
The structure of the Yukawa matrices $\tilde{Y}_1$ and $\tilde{Y}_2$ 
is constrained by the $U(1)'$ symmetry depending on its charge assignment. 
Throughout the paper, we take the Class-I assignment defined in Ref.~\cite{Nomura:2019dhw}, where 
$\Phi_1$ and the right-handed tau lepton $\tau_R^{}$ are charged with $q (\neq 0)$ and $-q$, respectively, 
while all the other fields are neutral.~\footnote{The same structure of the Yukawa interaction can be realized 
by imposing a $Z_2$ symmetry instead of the $U(1)'$ symmetry, where $\Phi_1$ and $\tau_R$ are assigned to be odd. 
In this case, an additional term $(\Phi_1^\dagger \Phi_2)^2$ appears in the Higgs potential.} 
This assignment provides the largest number of non-zero elements of Yukawa interaction matrices given as follows: 
\begin{align}
\tilde{Y}_1 = \begin{pmatrix}
0 & 0 & \times &0\\
0 & 0 & \times &0\\
0 & 0 & \times &0\\
0 & 0 & \times &0
\end{pmatrix},~~
\tilde{Y}_2 = \begin{pmatrix}
\times & \times & 0& 0\\
\times & \times & 0& 0\\
\times & \times & 0& 0\\
\times & \times & 0& 0
\end{pmatrix},\label{eq:lepton-mat}
\end{align}
where $\times$ denotes a non-zero complex value. The fourth column has to be zero due to the gauge invariance. 
The matrix $\tilde{F}$ is the anti-symmetric $4\times 4$ form, so that it is described by six independent parameters. 
We note that Yukawa interaction terms for quarks are the same form as those in the SM, 
where only $\Phi_2$ couples to quarks due to the $U(1)'$ symmetry. 
Thus, FCNCs do not appear in the quark sector at tree level. 

In order to separately write fermion mass terms and interaction terms, we introduce the Higgs basis~\cite{Georgi:1978ri,Donoghue:1978cj} defined as 
\begin{align}
\begin{pmatrix}
\Phi_1 \\
\Phi_2 
\end{pmatrix}
=\begin{pmatrix}
c_\beta & -s_\beta \\
s_\beta & c_\beta
\end{pmatrix}
\begin{pmatrix}
\Phi \\
\Phi' 
\end{pmatrix},
\end{align}
where we introduced shorthand notation for the trigonometric functions as
$s_X^{} = \sin X$ and $c_X^{} = \cos X$. 
In addition, we defined $\tan\beta = \langle \Phi_2^0 \rangle/\langle \Phi_1^0 \rangle$ and 
\begin{align}
\Phi = 
\begin{pmatrix}
G^+ \\
\frac{h_1' + v + iG^0 }{\sqrt{2}}
\end{pmatrix},\quad 
\Phi' = 
\begin{pmatrix}
H^+ \\
\frac{h_2'  + iA }{\sqrt{2}}
\end{pmatrix}. \label{eq:phi}
\end{align}
In Eq.~(\ref{eq:phi}), $G^\pm$ and $G^0$ are the Nambu-Goldstone (NG) bosons which are absorbed into the longitudinal components of $W^\pm$ and $Z$ bosons, respectively, while 
$H^\pm$, $h_{1,2}'$ and $A$ are physical charged, CP-even and CP-odd Higgs bosons, respectively. 
The VEV $v$ is related to the Fermi constant $G_F$ by $v = (\sqrt{2}G_F)^{-1/2} \simeq 246$ GeV. 
In general, $h_1'$ $(H^\pm)$ can mix with $h_2'$ ($S^\pm$). 
Their mass eigenstates are defined as 
\begin{align}
\begin{pmatrix}
h_1' \\
h_2'
\end{pmatrix}
&=\begin{pmatrix}
c_{\alpha -\beta} & -s_{\alpha -\beta} \\
s_{\alpha -\beta} & c_{\alpha -\beta}
\end{pmatrix}
\begin{pmatrix}
H\\
h
\end{pmatrix}, \quad 
\begin{pmatrix}
H^\pm \\
S^\pm
\end{pmatrix}
=\begin{pmatrix}
c_\chi & -s_\chi \\
s_\chi & c_\chi
\end{pmatrix}
\begin{pmatrix}
H_1^\pm\\
H_2^\pm
\end{pmatrix}, 
\end{align}
where the mixing angles $\alpha$ and $\chi$ are expressed in terms of the parameters in the Higgs potential, see Eq.~(\ref{eq:mixing}).  
We identify $h$ with the discovered Higgs boson with a mass of about 125 GeV.  

In the Higgs basis, Eq.~(\ref{eq:yukawa}) is rewritten as 
\begin{align}
\mathcal{L}_{\rm lep}  & = - M'\bar{L}^4L^4
- \left[\bar{L}_L^A\left(\tilde{Y}_{\Phi}^{AB} \Phi + \tilde{Y}_{\Phi'}^{AB} \Phi' \right)\ell_R^B  + \overline{L_L^{cA}}\tilde{F}^{AB}(i\tau_2)L_L^BS^+ + \text{h.c.}\right], 
\end{align}
where 
\begin{align}
\tilde{Y}_{\Phi} = \tilde{Y}_1 c_\beta + \tilde{Y}_2 s_\beta, \quad \tilde{Y}_{\Phi'} = - \tilde{Y}_1 s_\beta + \tilde{Y}_2 c_\beta . \label{eq:mass1}
\end{align}
The mass matrix for the charged leptons is expressed by 
\begin{align}
\tilde{M}_{\ell}   = \begin{pmatrix}
\frac{v}{\sqrt{2}}\tilde{Y}_\Phi^{ab} & 0 \\
\frac{v}{\sqrt{2}}\tilde{Y}_\Phi^{4b} & M'
\end{pmatrix},\quad (a,b=1,\dots, 3). 
\end{align}
This matrix can be diagonalized by the unitary rotations $L_L = V_LL_L'$ and $\ell_R = V_R\ell_R'$ such that 
\begin{align}
M_{\ell} \equiv \text{diag}(m_e,m_\mu,m_\tau,m_T) = V_L^\dagger \tilde{M}_\ell V_R. \label{eq:ml}
\end{align}
The interaction terms are then extracted in the mass eigenstates of the Higgs bosons and the charged fermions as 
\begin{align}
\mathcal{L}_{\rm lep}^{\rm int}  & =
- \frac{1}{\sqrt{2}}\bar{\ell}' \left[ (s_{\beta - \alpha} Y_{\Phi} + c_{\beta - \alpha} Y_{\Phi'})h + (c_{\beta - \alpha} Y_{\Phi} - s_{\beta - \alpha} Y_{\Phi'})H
+ i Y_{\Phi'}A  \right]P_R \ell'  + \text{h.c.} \notag\\
& -(2s_\chi \overline{\nu^{\prime c}}F P_L  + c_\chi\bar{\nu}' Y_{\Phi'} P_R    )\ell' \, H_1^+
  -(2c_\chi \overline{\nu^{\prime c }}F P_L  - s_\chi\bar{\nu}^{\prime } Y_{\Phi'} P_R    )\ell^{\prime } \, H_2^+  
   + \text{h.c.},  \label{eq:yuk} 
\end{align}
where $P_L$ ($P_R$) is the projection operator for the left (right) handed fermions, and 
\begin{align}
Y_{\Phi} &= V_L^\dagger \tilde{Y}_{\Phi} V_R,\quad Y_{\Phi'} = V_L^\dagger \tilde{Y}_{\Phi'} V_R, \quad F = V_L^T\tilde{F}V_L \label{ye}.  
\end{align}
These Yukawa matrices can be rewritten by using Eq.~(\ref{eq:ml}) as, 
\begin{align}
Y_{\Phi} = \frac{\sqrt{2}}{v}M_\ell V_R^\dagger P_{123} V_R, \quad 
Y_{\Phi'} = \frac{\sqrt{2}}{v}M_\ell V_R^\dagger P_\beta V_R,  \label{eq:ye2}
\end{align}
where $P_{123}\equiv \text{diag}(1,1,1,0)$ and $P_\beta \equiv \text{diag} (\cot\beta,\cot\beta,-\tan\beta,0)$. 
We note that $V_L$ does not appear in the above expression due to its unitarity.

In general, the matrices $Y_{\Phi}$ and $Y_{\Phi'}$ contain non-zero off-diagonal elements which 
can introduce large effects on CLFV decays mediated by scalar bosons. 
In order to avoid such CLFV decays, we assume that {\it the fourth lepton doublet $L^4$ is weakly mixed with  three generations of the SM leptons}. 
This can naturally be realized by taking the mass parameter $M'$ to be much larger than charged lepton masses, e.g., the tau lepton mass. 

Let us clarify how the interaction matrices defined above can be expressed in the weak mixing scenario. 
First, the unitary matrices $V_{R}$ can be expressed as 
\begin{align}
V_{R} &= V_{23}(\alpha_{23},\varphi_{23})V_{13}(\alpha_{13},\varphi_{13})V_{12}(\alpha_{12},\varphi_{12})
V_{14}(\alpha_{14},\varphi_{14})V_{24}(\alpha_{24},\varphi_{24})V_{34}(\alpha_{34},\varphi_{34}),  \label{eq:vr}
\end{align}
where $V_{ij}$ are the $4\times 4$ unitary matrices with $i$--$i$ and $j$--$j$ elements to be $\cos\alpha_{ij}$ and 
$i$--$j$ ($j$--$i$) to be $\sin\alpha_{ij}e^{-i\varphi_{ij}}$ ($-\sin\alpha_{ij}e^{i\varphi_{ij}}$), while all the other diagonal (off-diagonal) elements 
to be unity (zero). 
Second, the weak mixing scenario indicates that the mixing angles $\alpha_{a4}$ $(a=1,2,3)$ are small, so that 
we reparametrize them as $\alpha_{a4} \to \epsilon\alpha_{a4}$ with $\epsilon \ll 1$. 
In this case, the unitary matrices can be expressed  as, 
\begin{align}
V_R &= V_{R,0} + \delta V_R, 
\end{align}
where $V_{R,0}$ is given by taking $\epsilon \to 0$ in $V_R$, while $\delta V_{R}$ is 
\begin{align}
\delta V_{R} = 
\left(\begin{array}{c|c}
{\cal O}(\epsilon^2) & {\cal O}(\epsilon) \\ \hline 
{\cal O}(\epsilon) & {\cal O}(\epsilon^2)
\end{array}\right), 
\end{align}
with the upper-left (lower-right) block being the $3\times 3$ ($1\times 1$) form. 
Third, using this expansion, the Yukawa matrices $Y_{\Phi}$ and $Y_{\Phi'}$ are expressed as 
\begin{align}
Y_\Phi = Y_{\Phi,0} + \delta Y_\Phi,\quad Y_{\Phi'} = Y_{\Phi',0} + \delta Y_{\Phi'}, 
\end{align}
where 
\begin{align}
Y_{\Phi',0 } &= \frac{\sqrt{2}}{v}
\left(\begin{array}{c|c}
M_\ell^{3\times 3} (V_{R,0}^{3\times 3})^\dagger P_\beta^{3\times 3} V_{R,0}^{3\times 3} & 0 \\ \hline 
0 & 0
\end{array}\right), \label{eq:y0}   \\
\delta Y_{\Phi'} &= \frac{\sqrt{2}}{v}
\left(\begin{array}{c|c}
\epsilon^2 M_\ell^{3\times 3}[(\delta V_R^{3\times 3})^\dagger P_\beta^{3\times 3} V_{R,0}^{3\times 3} + \text{h.c.}]  & \epsilon M_\ell^{3\times 3}(V_{R,0}^{3\times 3})^\dagger  P_\beta^{3\times 3}\delta V_R^{3\times 1}\\ \hline 
\epsilon m_{T} (\delta V_R^{3\times 1})^\dagger  P_\beta^{3\times 3} V_{R,0}^{3\times 3 } & \epsilon^2 m_{T}(\delta V_R^{3\times 1})^\dagger P_\beta^{3\times 3}(\delta V_R^{3\times 1})
\end{array}\right)+ {\cal O}(\epsilon^3),
\end{align}
with $X^{3\times 3}$ ($X^{3\times 1}$) being the $3\times 3$ $(3\times 1)$ part of a $4 \times 4$ matrix $X$. 
The expressions for $Y_{\Phi,0}$ and $\delta Y_{\Phi}$ are given by replacing $P_\beta^{3\times 3} \to I^{3\times 3}$ in $Y_{\Phi',0}$ and $\delta Y_{\Phi'}$, respectively. 
We can see that the bottom-left block ($1\times 3$ part) of $\delta Y_\Phi$ and $\delta Y_{\Phi'}$ can be ${\cal O}(1)$ for $\epsilon \sim 1$ because of the dependence of $m_T$, 
and this effect can enter in the CLFV decays, e.g., $\ell \to \ell'\gamma$ decays via the fourth lepton loops. 
Thus, the weak mixing scenario, $\epsilon \ll 1$,  is essentially important to avoid such a large effect, as amplitudes for the CLFV decay are highly suppressed by $\epsilon^2$. 
Detailed discussions for constraints from the CLFV decays will be given in Sec.~\ref{sec:flavor}. 
Interestingly, the mixing effect on masses of active neutrinos is not suppressed by $\epsilon^2$, but $\epsilon$ as it will be explained in the next section. 
It is also important to mention here that $F$, $\delta Y_\Phi$ and $\delta Y_{\Phi'}$ can be complex, so that they
can provide new sources of CP-violation, and their effects on EDMs will be discussed in Sec.~\ref{sec:flavor}. 

It is clear that in the $\epsilon \to 0$ limit, $Y_{\Phi}$ becomes diagonal, while $Y_{\Phi'}$ takes the block diagonal form as shown in Eq.~(\ref{eq:y0}). 
Furthermore, in the scenario with a softly-broken $Z_2$ symmetry, the matrix $Y_{\Phi'}$ also becomes proportional to the diagonalized mass matrix as in the two Higgs doublet models (THDMs):  
\begin{align}
Y_{\Phi'} \xrightarrow[Z_2]{} \frac{\sqrt{2}M_\ell}{v}\zeta,
\end{align}
where $\zeta = \cot\beta$ ($-\tan\beta$) for the Type-I and Type-Y (Type-II and Type-X) THDM~\cite{Aoki:2009ha}.

\subsection{Higgs Potential}

The most general Higgs potential is given by 
\begin{align}
V  &= m_1^2|\Phi_1|^2+m_2^2|\Phi_2|^2  -m_3^2(\Phi_1^\dagger\Phi_2 +  \text{h.c.}) + m_S^2|S^+|^2 + \mu [\Phi_1^T (i\tau_2)\Phi_2 (S^+)^* + \text{h.c.}]\notag\\
& +\frac{\lambda_1}{2}|\Phi_1|^4 + \frac{\lambda_2}{2}|\Phi_2|^4+\lambda_3|\Phi_1|^2|\Phi_2|^2+\lambda_4|\Phi_1^\dagger\Phi_2|^2 \notag\\ 
& +\sigma_1|S^+|^2|\Phi_1|^2+\sigma_2|S^+|^2|\Phi_2|^2+\frac{\sigma_3}{2}|S^+|^4. 
\end{align}
In the above expression, $m_3^2$ and $\mu$ terms softly break the $U(1)'$ symmetry, and their complex phases 
are removed by using the phase redefinition of the scalar fields without the loss of generality. 
Thus, there is no CP-violating phase in the Higgs potential. 
We note that the $(\Phi_1^\dagger \Phi_2)^2$ term is forbidden because of the $U(1)'$ symmetry. 
Thus, a pseudo-NG boson, corresponding to $A$, appears due to the spontaneous breaking of the $U(1)'$ symmetry. 
The $\mu$ term plays a crucial role for the neutrino mass generation, as this term breaks two units of the lepton number when we assign the lepton number of $(\Phi_1,\Phi_2,S^\pm)$ to be $(0,0,\pm 2)$\footnote{By this assignment, 
the lepton number is conserved in the Yukawa interaction terms.  }. 

After imposing the tadpole conditions for two CP-even Higgs bosons, 
all the masses of the physical Higgs bosons are expressed as 
\begin{align}
m_A^2&=\frac{m_3^2}{s_\beta c_\beta}, \label{eq:masq}  \\
m_H^2&=(M_\text{even}^2)_{11}^2 \, c^2_{\beta-\alpha} + (M_\text{even}^2)_{22}^2\, s^2_{\beta-\alpha} - (M_\text{even}^2)_{12}^2\,s_{2(\beta-\alpha)},    \\
m_h^2&= (M_\text{even}^2)_{11}\,s^2_{\beta-\alpha} + (M_\text{even}^2)_{22}c^2_{\beta-\alpha} + (M_\text{even}^2)_{12}s_{2(\beta-\alpha)} , \\
m_{H_1^\pm}^2 & =  (M_\pm^2)_{11}\,c^2_{\chi} +(M_\pm^2)_{22}\, s^2_{\chi}  +  (M_\pm^2)_{12}\,s_{2\chi},    \\
m_{H_2^\pm}^2 & = (M_\pm^2)_{11} \,s^2_{\chi} +  (M_\pm^2)_{22}\,c^2_{\chi} -  (M_\pm^2)_{12}\,s_{2\chi} ,
\end{align}
where $(M_\text{even}^2)_{ij}$ and $(M_\pm^2)_{ij}$ ($i,j=1,2$) are the elements of the squared mass matrices 
for the CP-even and singly-charged Higgs bosons in the basis of $(h_1',h_2')$ and $(H^\pm, S^\pm)$, respectively. 
Each element is given as 
\begin{align}
(M_\text{even}^2)_{11}   & =v^2 \left(\lambda_1c^4_\beta+\lambda_2 s^4_\beta + \frac{\lambda_3+\lambda_4}{2} s^2_{2\beta} \right),\\
(M_\text{even}^2)_{22} & = m_A^2 + \frac{v^2}{8}[\lambda_1+\lambda_2-2(\lambda_3+\lambda_4)](1 - c_{4\beta}), \\
(M_\text{even}^2)_{12} & = \frac{v^2}{2} [-\lambda_1 c^2_\beta + \lambda_2 s^2_\beta + (\lambda_3+\lambda_4)c_{2\beta}]s_{2\beta},\\
(M_\pm^2)_{11} & = m_A^2-\frac{v^2}{2}\lambda_4, \\
(M_\pm^2)_{22} & = m_S^2+\frac{v^2}{2}(\sigma_1 c^2_\beta + \sigma_2 s^2_\beta), \\
(M_\pm^2)_{12} & = -v\frac{\mu}{\sqrt{2}}. 
\end{align}
The mixing angles are expressed in terms of these matrix elements: 
\begin{align}
\tan2(\alpha - \beta) &=\frac{2(M_{\text{even}}^2)_{12}}{(M_{\text{even}}^2)_{11} - (M_{\text{even}}^2)_{22}}, ~\tan2\chi=\frac{2(M_\pm^2)_{12}}{(M_\pm^2)_{11} - (M_\pm^2)_{22}}. \label{eq:mixing}
\end{align}

From the above discussion, we can choose the following twelve parameters as inputs:
\begin{align}
m_h,~
m_A,~ 
m_H,~
m_{H_1^\pm},~
m_{H_2^\pm},~
m_S,~
v,~
s_{\beta-\alpha},~
s_\chi,~
\tan\beta,~
\tan\sigma,~\sigma_3, 
\end{align}
where $\tan\sigma\equiv \sigma_2/\sigma_1$. 
Among the above parameters, $v$ and $m_h$ are fixed to be about 246 GeV and 125 GeV by experiments, respectively, and 
$\sigma_3$ is not relevant to the following discussion. 

The parameters of the Higgs potential are constrained by considering bounds from perturbative unitarity and vacuum stability. 
In Refs.~\cite{Muhlleitner:2016mzt,Chen:2020tfr}, all the independent eigenvalues of the $s$-wave amplitude matrix for 2 body to 2 body elastic scatterings  ($a_i$) have been given in the high energy limit. 
Requiring $|a_i| \leq 1/2$, the unitarity bound is expressed in our notation as 
\begin{align}
&\left|\frac{1}{2}\left(\lambda_{1}+\lambda_{2}+\sqrt{\left(\lambda_{1}-\lambda_{2}\right)^{2}+4 \lambda_{4}^{2}}\right) \right| <8\pi, \notag\\
& |\lambda_{1,2,3},\sigma_{1,2}|  <8\pi, \quad \left|\lambda_3 + 2 \lambda_4 \right|<8 \pi,\quad 
\left|\lambda_3 \pm \lambda_4\right| <8 \pi, \quad |x_{1,2,3}| <8 \pi, \label{eq:per-lam268}
\end{align}
where $x_{1,2,3}$ are the eigenvalues for the following $3\times 3$ matrix
\begin{align}
\begin{pmatrix}
3\lambda_1 & 2\lambda_3 + \lambda_4 & \sqrt{2}\sigma_1 \\
2\lambda_3 + \lambda_4 & 3\lambda_2 & \sqrt{2}\sigma_2\\
\sqrt{2}\sigma_1  & \sqrt{2}\sigma_2  & 2\sigma_3
\end{pmatrix}. 
\end{align}
The vacuum stability of the potential has also been discussed in Refs.~\cite{Kanemura:2000bq,Muhlleitner:2016mzt}. 
Requiring the potential bounded from below in any direction with large scalar field values, 
the parameters in the Higgs potential are constrained to be within the following domain~\cite{Muhlleitner:2016mzt}
\begin{equation}
\Omega_1\cup\Omega_2, 
\end{equation}
where 
\begin{align}
&
\Omega_{1}=\Big\{\lambda_{1}, \lambda_{2}, \sigma_{3} >0 ; \sqrt{\lambda_{1} \sigma_{3}}+ \sigma_{1}>0 ; \sqrt{\lambda_{2} \sigma_{3}}+\sigma_{2}>0 ; \notag \\
& \qquad\qquad
\sqrt{\lambda_{1} \lambda_{2}}+\lambda_{3}+D>0 ; \sigma_{1} +\sqrt{\frac{\lambda_{1}}{\lambda_{2}}} \sigma_{2} \geq 0\Big\},
\label{eq:stability1}
\\
&
\Omega_{2}=\Big\{\lambda_{1}, \lambda_{2},  \sigma_{3}>0 ; \sqrt{\lambda_{2} \sigma_{3}} \geq \sigma_{2}>-\sqrt{\lambda_{2} \sigma_{3}} ; \sqrt{\lambda_{1}\sigma_{3}}>-\sigma_{1}\geq 
\sqrt{\frac{\lambda_{1}}{\lambda_{2}}}\sigma_{2} ; \notag \\
& \qquad\qquad
\sqrt{\left(\sigma_{1}^{2}-\lambda_{1} \sigma_{3}\right)\left(\sigma_{2}^2 - \lambda_{2} \sigma_{3}\right)}> \sigma_{1}\sigma_{2} - \left(D+\lambda_{3}\right) \sigma_{3}\Big\}, 
\label{eq:stability2}
\end{align}
with $D = \min(0, \lambda_4)$. 

\section{Neutrino masses\label{sec:neutrino}}

Majorana masses for the active neutrinos are generated at one-loop level. 
In the $\nu_L'( = V_L\nu_L)$ basis, we obtain the mass matrix as
\begin{align}
m_\nu^{ab}  =  (FM_\ell C_{\rm diag} Y_{\Phi'}^\dagger)^{ab} + \text{transpose}, ~~~(a,b = 1,\dots,3)\label{eq:mnu}
\end{align}
where $C_{\rm diag}$ is the diagonal matrix given by   
\begin{align}
C_{\rm diag}^A = \frac{s_{2\chi}}{32\pi^2}\left[\ln  \frac{m_{H_2^\pm}}{m_{H_1^\pm}} + \frac{(m_\ell^A)^2 + m_{H_1^\pm}^2}{(m_\ell^A)^2 - m_{H_1^\pm}^2}\ln\frac{m_{H_1^\pm}}{m_\ell^A} 
- \frac{(m_\ell^A)^2 + m_{H_2^\pm}^2}{(m_\ell^A)^2 - m_{H_2^\pm}^2}\ln\frac{m_{H_2^\pm}}{m_\ell^A}  \right],  \label{eq:mn}
\end{align}
with $m_\ell^A$ being the charged lepton mass. For $m_\ell^A \ll m_{H_1^\pm},m_{H_2^\pm}$, 
\begin{align}
C_{\rm diag}^A \simeq \frac{s_{2\chi}}{32\pi^2}\ln \frac{m_{H_2^\pm}^2}{m_{H_1^\pm}^2} =: C_{\rm diag}^0. 
\end{align}
We note that the mass of the sterile neutrino $\nu_{T}$ is simply given by the vector-like mass $M'$.\footnote{Exactly speaking, 
the masses of the active neutrinos given in Eq.~(\ref{eq:mn}) are corrected by $m_\nu^2/M'$ from the mixing between $\nu_{T}$ and the SM three neutrinos, which is negligibly small for TeV scale $M'$. }

In the weak mixing scenario, the neutrino mass matrix can be expressed as 
\begin{align}
m_\nu  =  m_\nu^0 + \epsilon m_\nu^1 + \epsilon^2 m_\nu^2 + {\cal O}(\epsilon^3) ,  \label{eq:nu}
\end{align}
where 
\begin{align}
m_\nu^0  &= C_{\rm diag}^0 F^{3\times 3} M_\ell^{3\times 3} (Y_{\Phi',0}^{3\times 3})^\dagger+ \text{transpose}, \label{eq:nu0}\\
m_\nu^1  &=  m_{T}C_{\rm diag}^4  F^{3\times 1}  (\delta Y_{\Phi'}^{3\times 1}|_{\epsilon\to 1})^\dagger+ \text{transpose}, \label{eq:nu1}\\
m_\nu^2  &=  C_{\rm diag}^0 F^{3\times 3} M_\ell^{3\times 3} (\delta Y_{\Phi'}^{3\times 3}|_{\epsilon\to 1})^\dagger+ \text{transpose}.\label{eq:nu2} 
\end{align}
We note that the contribution from the order $\epsilon^2$ term $m_\nu^2$ is much smaller than that from $m_\nu^1$ due to 
not only the suppression of the $\epsilon$ factor but also no enhancement by the large lepton mass $m_{T}$. Therefore, 
the neutrino masses can be well approximated by considering the term up to $m_\nu^1$. 
The neutrino mass matrix becomes the one given in the original Zee model by taking $\epsilon \to 0$ and $P_\beta^{3\times 3} \to \cot\beta I^{3\times 3}$ 
in $Y_{\Phi',0}$, which is consistent with the expression given in Ref.~\cite{He:2003ih}; 
\begin{align}
m_\nu \to m_\nu^0 =  
\frac{\sqrt{2}s_{2\chi}}{32\pi^2 v}\cot\beta \ln\left( \frac{m_{H_2^\pm}^2}{m_{H_1^\pm}^2}\right) [F^{3\times 3}(M_\ell^{3\times 3})^2 - (M_\ell^{3\times 3})^2F^{3\times 3} ].   \label{eq:mnu2_xxx} 
\end{align}
As aforementioned, this structure cannot accommodate current neutrino oscillation data. 

The neutrino mass matrix can be diagonalized by introducing the PMNS matrix $U_{\rm PMNS}$ as follows: 
\begin{align}
U^T_{\rm PMNS}\, m_\nu \, U_{\rm PMNS} = m_\nu^{\rm diag} = \text{diag}(m_1,m_2,m_3), \label{eq:nudat} 
\end{align}
where $m_i$ ($i=1,2,3$) are the mass eigenvalues. For the normal ordering (NO) and the inverted ordering (IO) cases, these are $m_1 \lesssim m_2 \ll m_3$ and 
$m_3 \ll m_1 \lesssim m_2$, respectively. 
The matrix $U_{\rm PMNS}$ can be parameterized as, 
\begin{align}
&U_{\rm PMNS} \equiv \begin{pmatrix}
1 &0& 0 \\
0 &c_{\theta_{23}} & s_{\theta_{23}} \\
0 &-s_{\theta_{23}} & c_{\theta_{23}}
\end{pmatrix}
\begin{pmatrix}
c_{\theta_{13}} &0& s_{\theta_{13}}e^{-i\delta_{\rm CPV}} \\
0 &1& 0 \\
-s_{\theta_{13}}e^{i\delta_{\rm CPV}} &0& c_{\theta_{13}}
\end{pmatrix}
\begin{pmatrix}
c_{\theta_{12}}  & s_{\theta_{12}} & 0 \\
-s_{\theta_{12}} & c_{\theta_{12}}& 0 \\
0 &0& 1
\end{pmatrix}
\begin{pmatrix}
e^{i\varphi_1} & 0 & 0 \\
0 &e^{i\varphi_2} & 0 \\
0 &0& 1
\end{pmatrix},  \label{eq:uell}
\end{align}
where $\delta_{\rm CPV}$ is the Dirac CP-phase, and $\varphi_{1,2}$ are the Majorana phases. 
From Eqs.~(\ref{eq:nu}) and (\ref{eq:nudat}), 
we obtain 
\begin{align}
m_\nu \simeq m_\nu^0 + \epsilon m_\nu^1 = U_{\rm PMNS}^* \, m_\nu^{\rm diag} \, U_{\rm PMNS}^\dagger. \label{eq:dat}
\end{align}
By solving this equation, the elements of $F$ are expressed in terms of the following model parameters: 
\begin{align}
\alpha_{AB},~~\varphi_{AB},~~\tan\beta,~~\epsilon,~~m_{T},~~m_{H_{1,2}^\pm},~~s_\chi, \label{eq:nu-theory}
\end{align}
with $\{\alpha_{AB}\} = \{\alpha_{12},~\alpha_{13},~\alpha_{23},~\alpha_{14},~\alpha_{24},~\alpha_{34}\}$ (similarly to $\varphi_{AB}$), 
and the neutrino parameters appearing in the right-hand side of Eq.~(\ref{eq:dat}):
\begin{align}
\theta_{12},~~\theta_{23},~~\theta_{13},~~\Delta m_{\text{sol}}^2,~~|\Delta m_{\text{atm}}^2|,~~m_{0},~~\delta_{\rm CPV},~~\varphi_{1,2}, \label{eq:nu-dat}
\end{align}
with $m_{0}$ being the smallest eigenvalue of the neutrino masses. 
We define the two squared mass differences as $\Delta m_{\rm sol}^2 \equiv m_2^2 - m_1^2$ and $\Delta m_{\rm atm}^2 \equiv |m_3^2 - m_1^2|~~(|m_3^2 - m_2^2|)$ 
for the NO (IO) case. 

Now, we can reproduce the neutrino parameters shown in Eq.~(\ref{eq:nu-dat}) for each fixed value of the model parameters given in Eq.~(\ref{eq:nu-theory}). 
Typical order of the elements of $F$ can be estimated for $\tan\beta \simeq 1$ as follows: 
\begin{align}
F^{AB} \simeq \frac{1}{C_{\rm diag}^B}\frac{v}{m_\ell^A}\frac{m_\nu}{m_\ell^B}[1 + (\epsilon^{-1} -1)\delta^{4B}]~~(\text{for}~~A<B), 
\end{align}
where $m_\nu$ is the typical value of the elements of the neutrino mass matrix, i.e., $m_\nu = {\cal O}(10^{-3}\text{--}10^{-2})$ eV.
In Appendix~\ref{sec:fmat}, we present better approximated formulae for these elements.

Let us comment on the case without the mixing between $L^4$ and the SM leptons, i.e., $\epsilon = 0$ (or equivalently taking $M' \to \infty$). 
In this case, we have $C_{\rm diag}^0$, $\tan\beta$, $F^{ab}$, $\alpha_{ab}$ and $\varphi_{ab}$ with $(a,b)=(1,2),~(2,3)$ and $(1,3)$ as the free input parameters, among which 
$C_{\rm diag}^0$ can be chosen to satisfy one of the measured values of the squared mass differences. 
By scanning the parameters within the following regions: 
$0 \leq |F^{ab}| \leq 1$, $-\pi \leq \{\text{Arg}[F^{ab}],~\alpha_{ab},~\varphi_{ab}\} \leq \pi$ and $1 \leq \tan\beta \leq 10$, 
we obtain $27^\circ\lesssim \theta_{13}\lesssim 35^\circ$ and $2.6 \times 10^{-4}$ eV$^2 \lesssim \Delta m_{21}^2 \lesssim 2.5 \times 10^{-3}$ eV$^2$ in the NO case
under the requirement that $\theta_{12}$, $\theta_{23}$ and $|\Delta m_{\rm atm}^2|$ are given within the $3\sigma$ region of the global fit results~\cite{Esteban:2020cvm,Nufit}. 
For the IO case, we obtain $40^\circ\lesssim \theta_{12}\lesssim 45^\circ$ and $1.5 \times 10^{-4}$ eV$^2\lesssim \Delta m_{21}^2 \lesssim 1.6 \times 10^{-3}$ eV$^2$
under the requirement that $\theta_{23}$, $\theta_{13}$ and $|\Delta m_{\rm atm}^2|$ are given within the $3\sigma$ region of the global fit results.~\cite{Esteban:2020cvm,Nufit}. 
Therefore, we clarify that the minimal model without the vector-like lepton cannot accommodate the current neutrino data.\footnote{As mentioned in Introduction, if we do not impose any new symmetries such as $Z_2$ and $U(1)$, 
we can find a solution to satisfy the current neutrino data. This, however, requires a fine-tuning in the quark Yukawa sector in order to suppress FCNCs mediated by scalar bosons. }

\section{Flavor Constraints\label{sec:flavor}}

\subsection{Electric Dipole Moments}

As we have discussed in the previous section, non-zero CP-violating phases appear in the lepton sector.
In this subsection, we study effects of these phases on EDMs, particularly the electron EDM\footnote{The CP-violating source 
in the quark Yukawa sector is the same as in the SM, i.e, only arising from the Kobayashi-Maskawa phase. Thus, its effect on EDMs 
is negligibly small~\cite{Pospelov:2005pr}. }. 
At one-loop level, contributions from the neutral Higgs bosons ($h,~H$ and $A$) are expressed as 
\begin{align}
d_e = -\frac{e}{32\pi^2}\sum_{A =1,4}\sum_{\phi = h,H,A}\frac{m_\ell^A}{m_\phi^2}\text{Im}[(Y_{\phi})_{1A}(Y_{\phi})_{A1}] A\left[\frac{(m_\ell^A)^2}{m_\phi^2}\right], \label{eq:edm}
\end{align}
where 
\begin{align}
Y_h = s_{\beta - \alpha} Y_{\Phi} + c_{\beta - \alpha} Y_{\Phi'},~ 
Y_H = c_{\beta - \alpha} Y_{\Phi} - s_{\beta - \alpha} Y_{\Phi'},~
Y_A = i Y_{\Phi'},  
\end{align}
and the loop function $A$ is given by 
\begin{align}
A(r) = \frac{1}{2(1-r)^2}\left(3 - r + \frac{2}{1-r}\ln r \right). 
\end{align}
From Eq.~(\ref{eq:ye2}), 
we can easily show that the product $(Y_{\Phi}Y_{\Phi})_{11}$ becomes a real value: 
\begin{align}
 (Y_{\Phi})_{1A}(Y_{\Phi})_{A1} = \frac{2m_e}{v^2}(V_R^\dagger P_{123}V_R)^{1A}m_\ell^A(V_R^\dagger P_{123}V_R)^{A1} = 
\frac{2m_e}{v^2}(|V_R^\dagger P_{123}V_R|^{2})^{1A}m_\ell^A. \label{eq:edm2_1}
\end{align}
A similar argument holds for $(Y_{\Phi'}Y_{\Phi'})_{11}$ just by replacing $P_{123} \to P_{\beta}$ in the above expression. 
On the other hand, a non-zero imaginary part appears from the product $(Y_{\Phi}Y_{\Phi'})_{11}$ as follows
\begin{align}
\text{Im}[(Y_{\Phi})_{1A}(Y_{\Phi'})_{A1}] = \frac{2m_e}{v^2}\text{Im}[(V_R^\dagger P_{123}V_R)^{1A}m_\ell^A(V_R^\dagger P_\beta V_R)^{A1}] \sim 
\frac{2m_em_T}{v^2}\epsilon^2.  
\end{align}
This combination appears from the $h$ or $H$ loop contribution, which is proportional to $s_{\beta-\alpha}c_{\beta-\alpha}$. 
Thus, the magnitude of the electron EDM can be estimated for $m_{T} \gg m_\phi$ as 
\begin{align}
\frac{d_e}{e} \sim \frac{\epsilon^2}{8\pi^2}s_{\beta-\alpha}c_{\beta-\alpha}\frac{m_e}{v^2} \sim 2.1\times 10^{-33}\times \left(\frac{\epsilon}{10^{-4}}\right)^2\times \left(\frac{s_{\beta-\alpha}c_{\beta-\alpha}}{0.1} \right)\text{cm}. \label{eq:edm2_2}
\end{align}
Therefore, the typical value of the electron EDM is much smaller than the current upper limit given from the ACME collaboration, i.e., $|d_e/e| < 1.1 \times 10^{-29}$ cm at the 90\% confidence level~\cite{Andreev:2018ayy}. 
We note that diagrams with charged Higgs boson loops are negligibly small, because these contributions are proportional to the neutrino mass\footnote{Contributions from $\nu_T$ and the charged Higgs boson loop
vanish, because these are proportional to \text{Im}($Y_{\Phi'}Y_{\Phi'})_{11}$.
}. 
Thus, the one-loop contributions to the electron EDM can safely be ignored. 

One might think that two-loop Barr-Zee (BZ) type diagrams~\cite{Barr:1990vd} 
give an important contribution to the electron EDM, rather than the one-loop contribution as in CP-violating THDMs~\cite{Jung:2013hka,Abe:2013qla,Kanemura:2020ibp}. 
However, their contributions also do not give a significant contribution as explained below. 
For BZ diagrams with a neutral gauge boson ($\gamma$ or $Z$) exchange, they are proportional to $\text{Im}(Y_{\Phi})_{11}$ or $\text{Im}(Y_{\Phi'})_{11}$ which is zero in our model. 
For those with a $W$ boson exchange, its contribution can further be separated into that with $H^\pm$ and $S^\pm$ exchanges, where 
the former is again proportional to $\text{Im}(Y_{\Phi'})_{11}$ and the latter is negligibly small as it picks us a tiny neutrino mass from an
internal neutrino line.\footnote{This argument is not qualitatively changed in the discussion with the mass eigenstates of the charged Higgs bosons $H_1^\pm$ and $H_2^\pm$. } 
Therefore, our model is safe from the constraints of the EDMs even if we consider ${\cal O}(1)$ CP-phases in the Yukawa couplings. 

\subsection{Constraints from Charged Lepton Flavor Violation}

In this subsection, we discuss CLFV processes.
The Yukawa interactions in Eq.~\eqref{eq:yuk} induce CLFV processes $\ell_i \to \ell_j \gamma (m_{\ell_i} > m_{\ell_j})$ where $\{\ell_1, \ell_2, \ell_3 \} = \{e, \mu, \tau \}$. 
Branching ratios (BRs) of these processes are given by
\begin{align}
&{\rm BR}(\ell_i\to\ell_j\gamma)\simeq \frac{48\pi^3\alpha_{\rm em}C_{ij}}{G_F^2 m^2_{\ell_i}} \left(\Big|\sum_\phi(a_R^\phi)_{ij}\Big|^2 + \Big|\sum_\phi(a_L^\phi)_{ij}\Big|^2\right), \label{br_clfv}
\end{align}
where $\alpha_{\rm em}$ is the fine structure constant, and 
$C_{ij}$ are numerical constants associated with the BR of the charged lepton, i.e.,  
$C_{21}= {\rm BR}(\mu \to e \nu_\mu \bar \nu_e) = 1$, $C_{31}= {\rm BR}(\tau \to e \nu_\tau \bar \nu_e)=0.1784$ and $C_{32} = {\rm BR}(\tau \to \mu \bar \nu_\mu \nu_e) =0.1736$. 
In Eq.~(\ref{br_clfv}), $a^\phi_{L,R}$ denote an amplitude which is estimated from a one-loop diagram with a scalar boson $\phi = \{ h, H, A, H^\pm_1, H^\pm_2 \}$ running in the loop;  
explicit forms of these amplitudes are summarized in Appendix~\ref{sec:LFV_amp}.

In addition we consider spin-independent $\mu \to e$ conversion via $H$ exchange; CP-odd scalar exchange induces spin-dependent $\mu \to e$ conversion process which is less constrained.
We can write the BR for the process such that~\cite{Kuno:1999jp,Kitano:2002mt,Davidson:2018kud}
\begin{align}
& BR(\mu \to e) = \frac{32 G_F^2 m_\mu^5}{\Gamma_{cap}} \left| C_{SL}^{pp} S^{(p)} + C_{SL}^{nn} S^{(n)} + C_{SR}^{pp} S^{(p)} + C_{SR}^{nn} S^{(n)} \right|^2, \\
& C_{SL[SR]}^{pp} \simeq C_{SL[SR]}^{nn} \simeq \frac{f_N m_N}{2 G_F m_H^2 v } \cot \beta (Y_{\Phi'})_{12[21]},
\end{align}
where $\Gamma_{cap}$ is the rate for the muon to transform to a neutrino by capture on the nucleus, $S^{(p,n)}$ is the integral over the nucleus for lepton wavefunctions with corresponding nucleon density, 
and $f_N \sim 0.3$ is effective coupling between Higgs and nucleon $N$ defined by $f_N m_N \bar N N = \sum_q m_q \langle N | \bar q q |  N \rangle $ with nucleon mass $m_N$~\cite{Cline:2013gha}.
The values of $\Gamma_{cap}$ and $S^{(n,p)}$ depend on target nucleus, and corresponding values for $^{197}_{79}$Au and $^{27}_{13}$Al targets  are obtained as 
$\Gamma_{cap}(^{197}_{79}{\rm Au} [^{27}_{13}{\rm Al}] ) = 13.07[0.7054] \times 10^6 {\rm sec}^{-1}$,  $S^{(p)}(^{197}_{79} {\rm Au} [^{27}_{13}{\rm Al}] ) = 0.0614[0.0155]$ and 
$S^{(n)}(^{197}_{79} {\rm Au} [^{27}_{13}{\rm Al}] ) = 0.0981[0.0167]$~\cite{Suzuki:1987jf,Kitano:2002mt}.

The current upper limits on the above BRs with 90\% confidence level are found in Refs.~\cite{TheMEG:2016wtm, Aubert:2009ag,Renga:2018fpd,Lindner:2016bgg,Bertl:2006up,Coy:2018bxr} as
\begin{align}
&{\rm BR}(\mu\to e\gamma) < 4.2\times10^{-13},~ 
{\rm BR}(\tau\to e\gamma) < 3.3\times10^{-8},~
{\rm BR}(\tau\to\mu\gamma) < 4.4\times10^{-8}, \nonumber \\
& {\rm BR}(\mu \to e)_{\rm Al} < 7 \times 10^{-13}. \label{eq:lfvs-cond}
\end{align}
We note that the Yukawa interactions also induce three body decays such as $\mu \to e e \bar e$ via the diagram with off-shell $H$ and $A$ at tree level.
BRs of such decay modes are found to be much smaller than the current upper bound, e.g. BR$(\mu \to ee \bar e) < 1.0 \times 10^{-12}$~\cite{Bertl:2006up}, because 
BR$(\mu \to ee \bar e)$ is proportional to the small lepton masses, i.e., $(m_em_\mu/v^2)^2 = {\cal O}(10^{-17})$. 
We also note that the Z boson couplings with right-handed charged leptons can contain flavor violation, because $\ell_R^{1\text{--}3}$ and $\ell_R^{4} (=T_R^4)$
belong to the different representation under the $SU(2)_L\times U(1)_Y$ group. 
In the mass eigenbasis, the coefficients of the $Z^\mu \bar{\ell}_R^{\prime a}\gamma_\mu\ell_R^{\prime b}$ vertices ($a,b=1,2,3$ and $a\neq b$) are proportional to $\epsilon^2$, so that 
the off-shell Z boson contribution to the decay rate of $\mu \to ee\bar{e}$  is suppressed by $\epsilon^4$.  
Thus, we can safely avoid the bound by taking $\epsilon < {\cal O}(10^{-3})$. \footnote{Milder bounds can be obtained from the decay of the $Z$ boson into a pair of different lepton flavor. 
The current upper limits on BR($Z \to \ell\ell')$ are given to be of order $10^{-7}$, $10^{-6}$ and $10^{-5}$ for the $e\mu$, $e\tau$ and $\mu\tau$ final state~\cite{Zyla:2020zbs}, respectively. 
We also check that bounds from the $\mu$-$e$ conversion provide a slightly stronger limit on $\epsilon$ as compared with those from $\mu \to 3 e$~\cite{Bertl:2006up,Crivellin:2020ebi,Crivellin:2020klg}.}
We thus concentrate on the constraints from the $\ell_i \to \ell_j \gamma$ processes in the following discussion. 

\begin{figure}[t!]\begin{center}
\includegraphics[width=60mm]{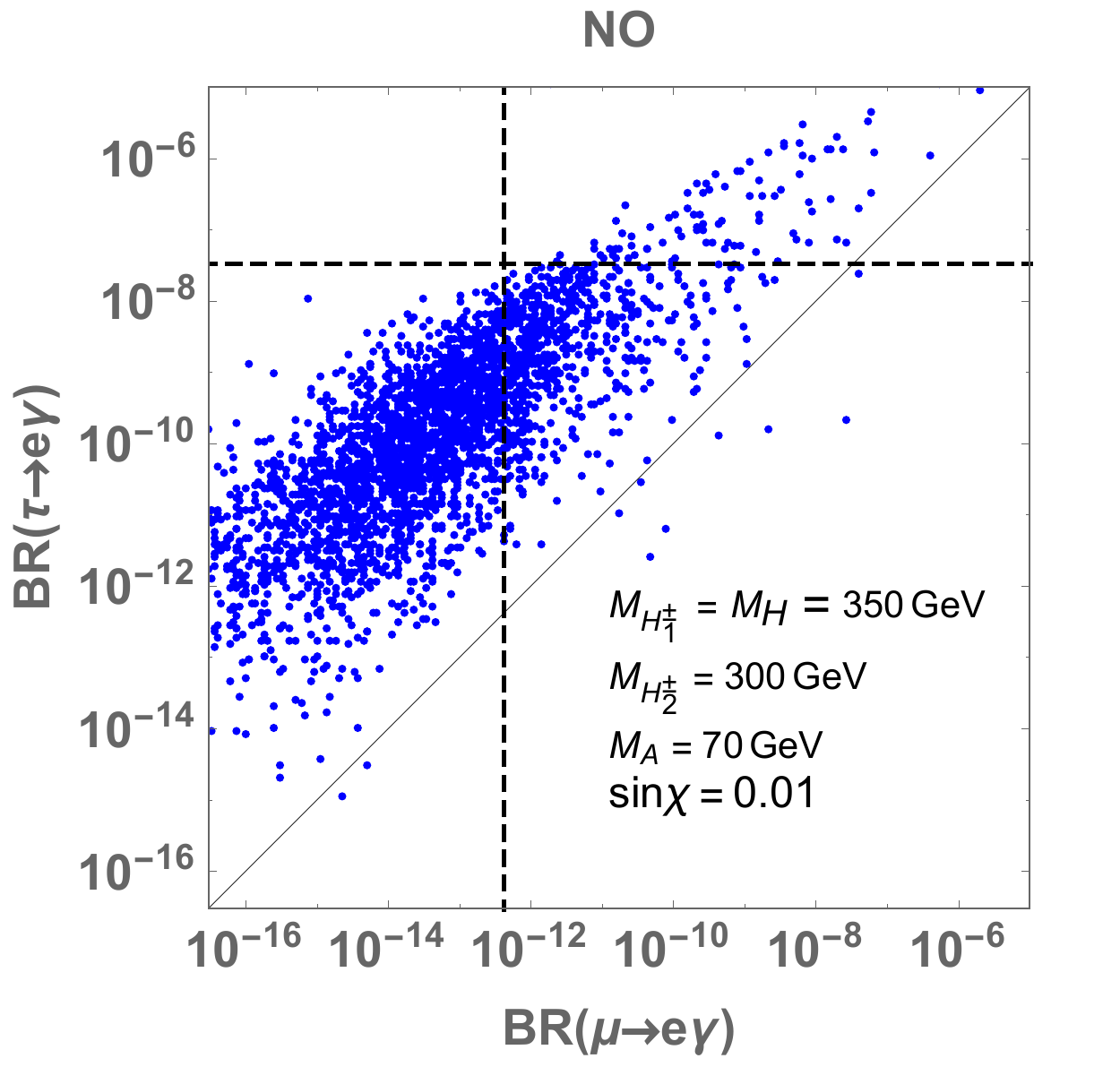} \qquad
\includegraphics[width=60mm]{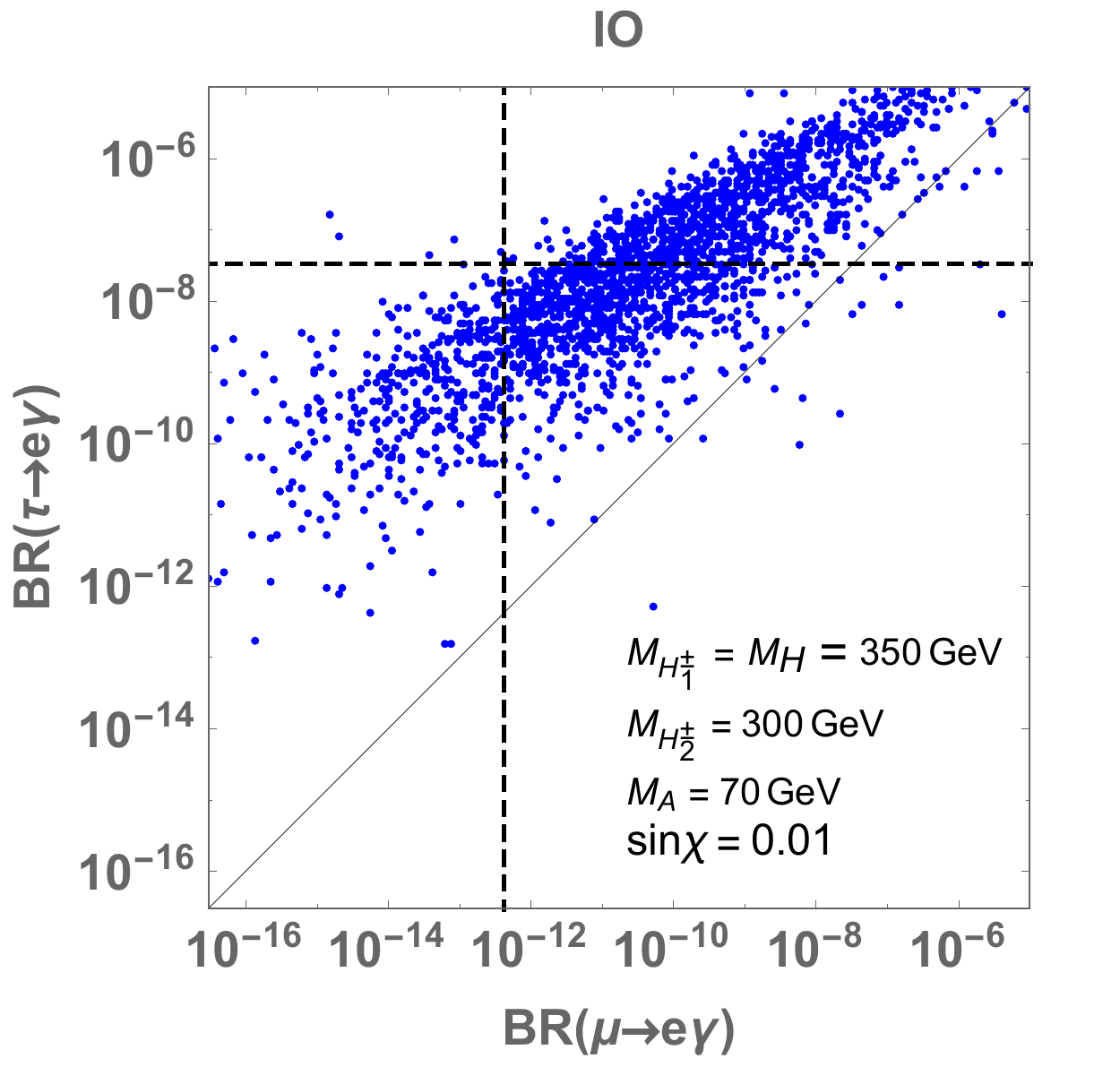} 
\includegraphics[width=60mm]{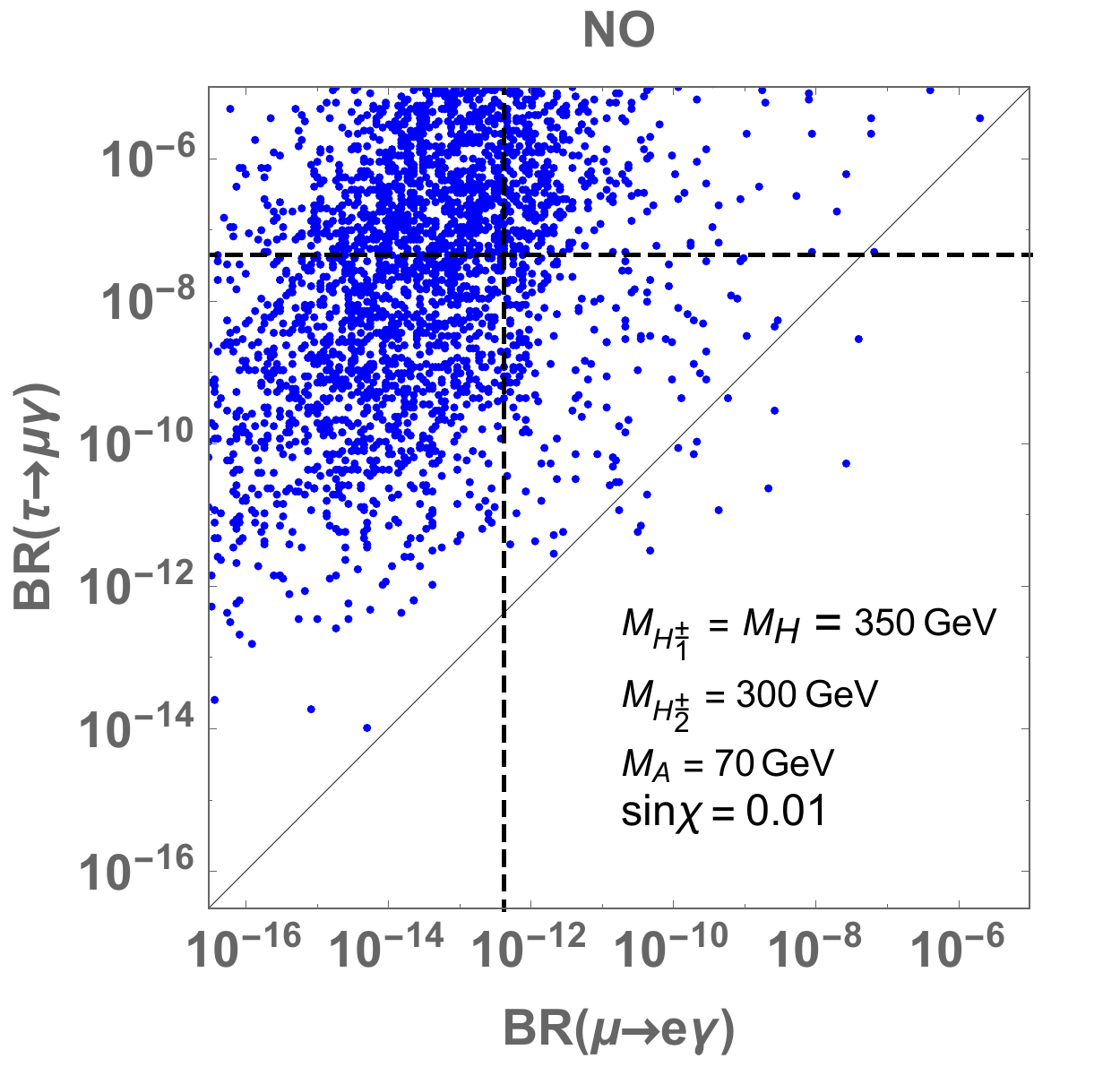} \qquad
\includegraphics[width=60mm]{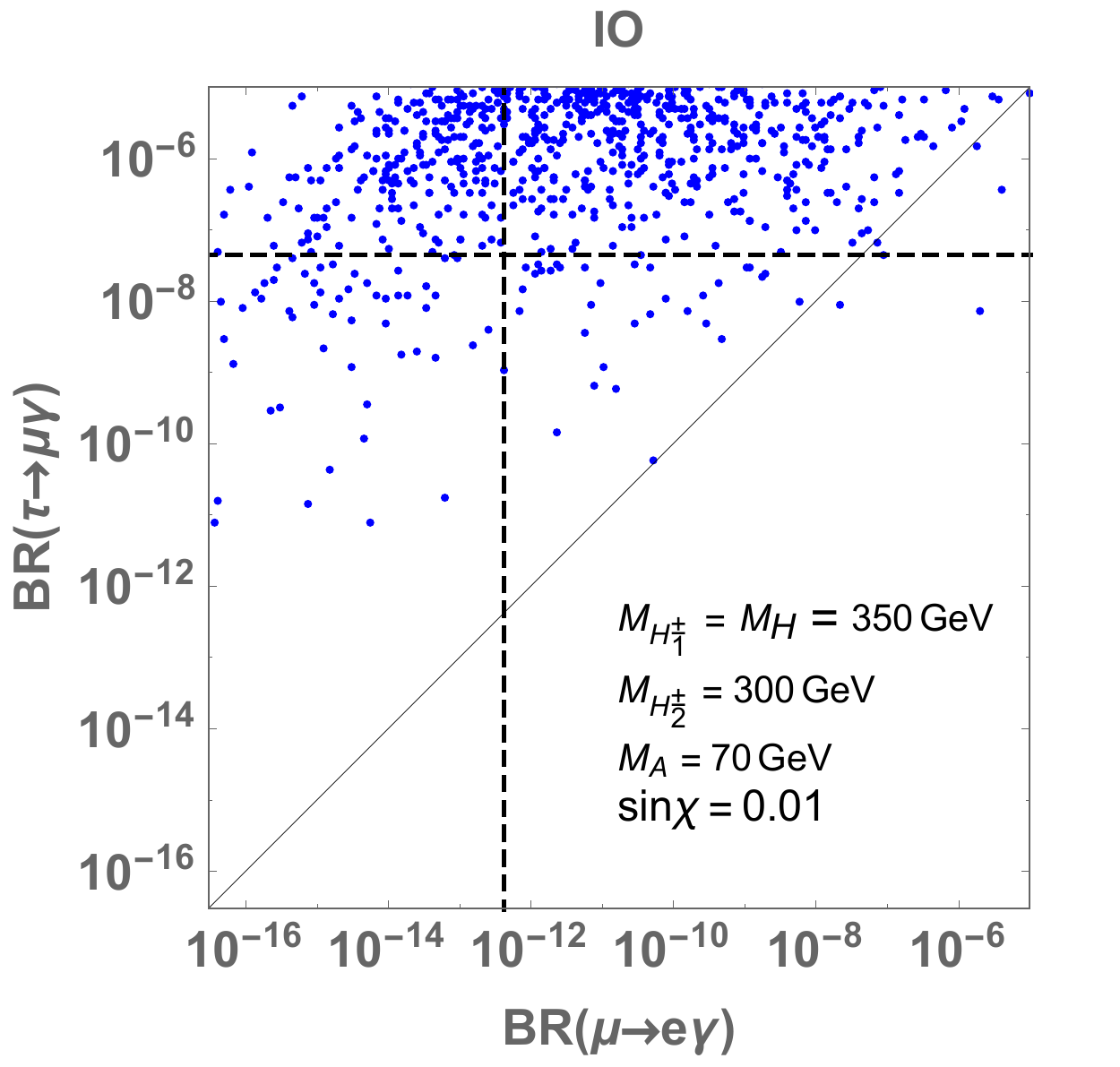}
\includegraphics[width=60mm]{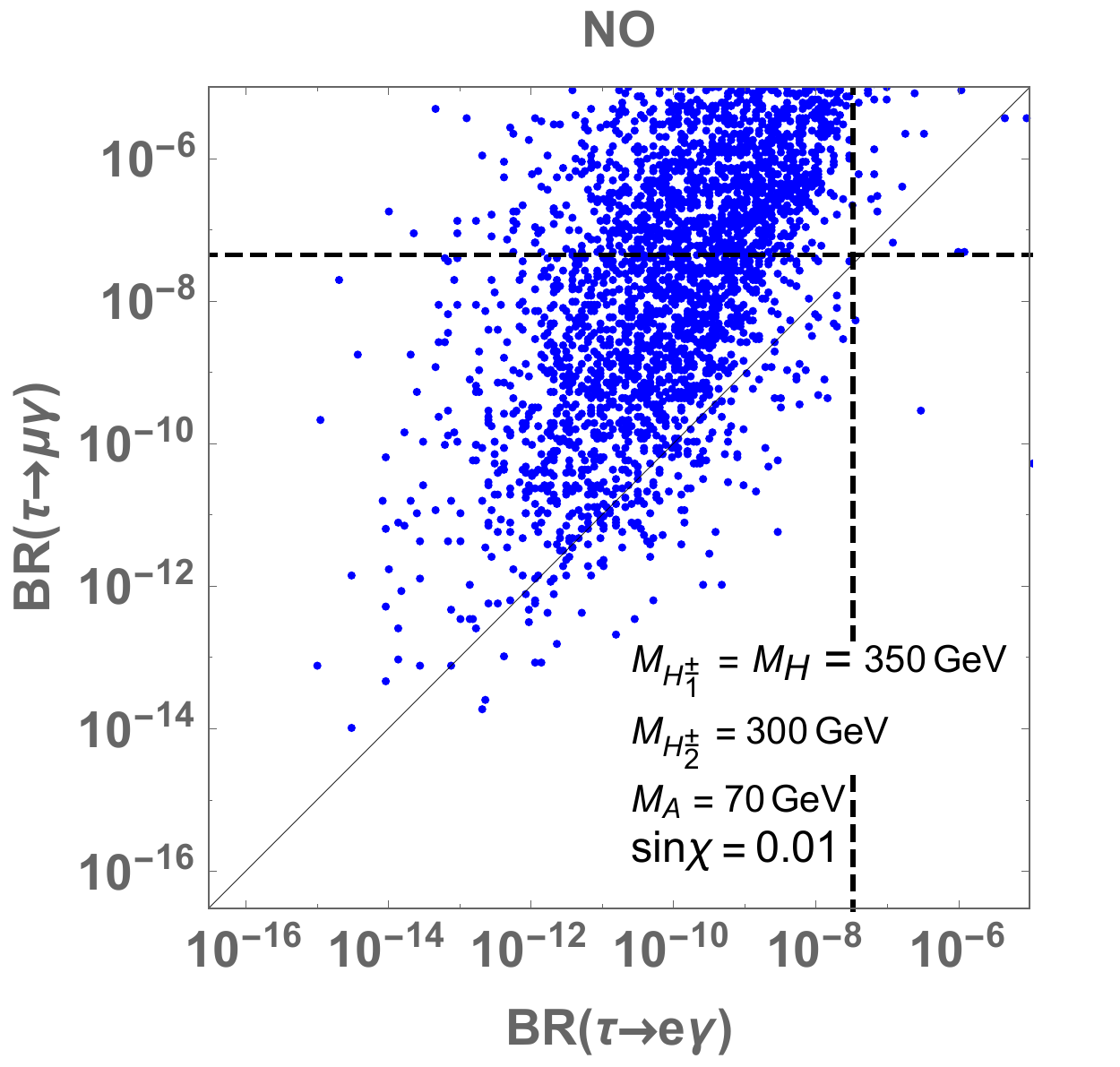} \qquad
\includegraphics[width=60mm]{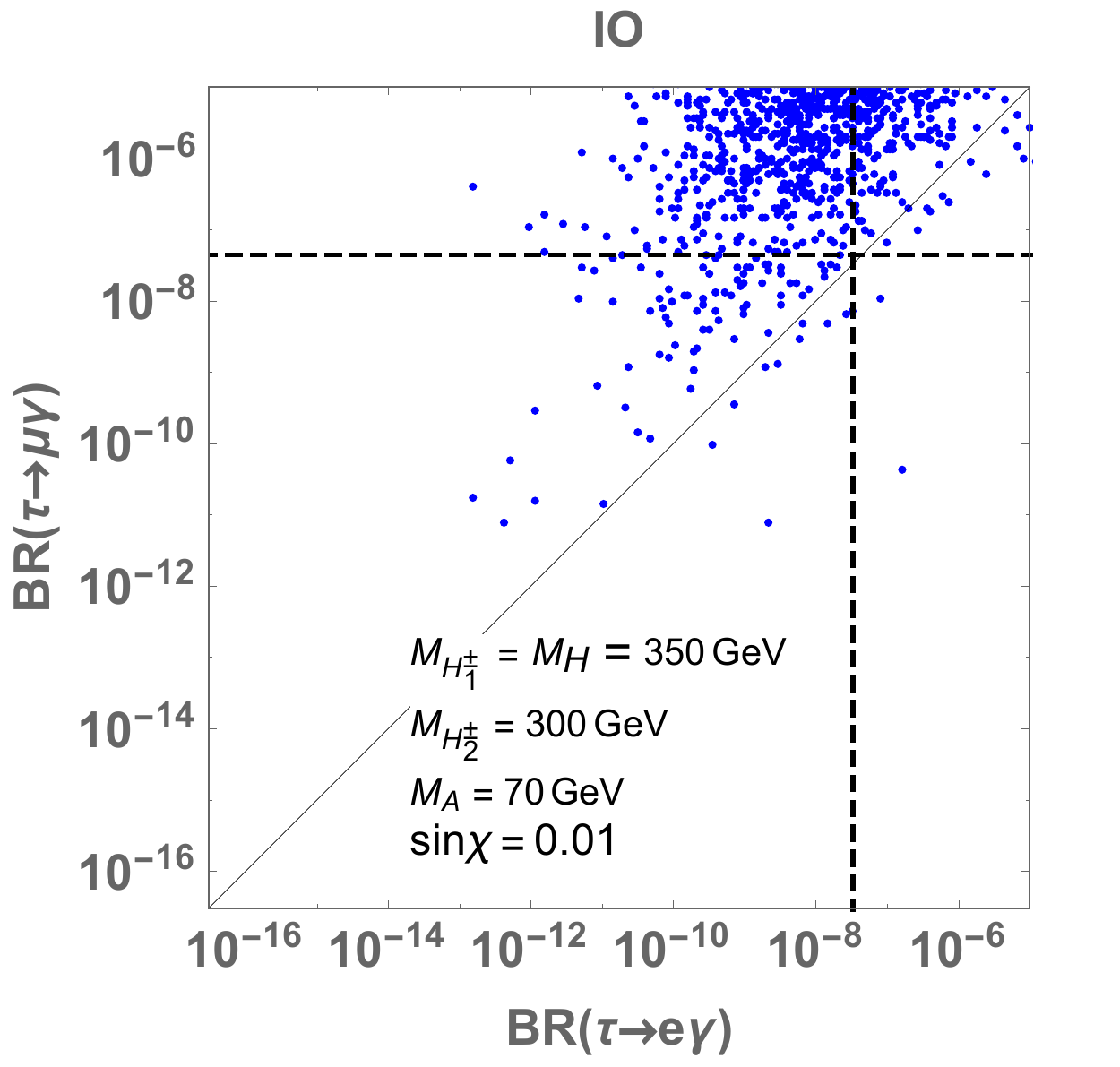}
\caption{Correlations among BR$(\ell_i \to \ell_j \gamma)$ for the case with 
$m_{H_1^\pm}^{} = m_H = 350$ GeV, $m_{H_2^\pm}^{} = 300$ GeV, $m_A = 70$ GeV where left(right) plots correspond to NO(IO).
The dashed horizontal and vertical lines indicate the upper bounds for each BR with 90\% confidence level. }   
\label{fig:LFV1}
\end{center}
\end{figure}

In the following, we perform numerical evaluations of the neutrino mass matrix and the BRs of the CLFV decays. 
First, we scan the input parameters as 
\begin{align}
 -\pi \leq \{\alpha_{AB}, \delta_{\rm CPV} \} \leq \pi,\quad 1 \leq \tan\beta \leq 10,\quad 10^{-5} \leq m_0 \leq 0.1~\text{eV}. \label{eq:inputs}
\end{align}
We fix the mass parameters to be $(m_{T},m_H,m_{H^\pm_1},m_{H_2^\pm},m_A) = (1000,350,350,300,70)$ GeV, where 
the scalar boson masses are chosen such that the FOPT can be realized as we see in the next section. 
In addition, the other parameters are fixed to be $s_\chi = 0.01$, $\epsilon = 10^{-4}$ and $\varphi_{1,2}=\varphi_{AB} =0$. 
We then solve Eq.~\eqref{eq:dat} to obtain $F^{AB}$ for each set of the input parameters, 
where we use the best fit values of the neutrino parameters $\theta_{ij}$, $|\Delta m_{\rm  atom}^2|$ and $\Delta m_{\rm sol}^2$ from the global fit results~\cite{Esteban:2020cvm,Nufit} for both the NO and IO cases. 
Finally, we apply the alignment limit $s_{\beta - \alpha} = 1$, which is supported by the current LHC data~\cite{Aad:2019mbh,Sirunyan:2018koj}, and
is favored by the constraint from the electroweak $\rho$ parameter in the case of $m_H^{} = m_{H_1^\pm}^{}$~\footnote{In our numerical calculation, 
we take the small mixing angle for the charged Higgs bosons, i.e., $s_\chi = 10^{-2}$, so that the $H_1^\pm$ states almost coincide with the charged Higgs bosons from the doublet $H^\pm$.  
Thus, the one-loop corrections to the $\rho$ parameter from the scalar boson loops almost completely vanish due to the approximate custodial symmetry~\cite{Pomarol:1993mu,Aiko:2020atr}. }. 

In Fig.~\ref{fig:LFV1}, we show the correlations between the BRs of $\ell_i \to \ell_j \gamma$ processes, where the left and right plots correspond to the NO and IO cases. 
The upper bounds of the BRs are indicated by the dashed lines.
We find a strong prediction of $\text{BR}(\tau \to \mu \gamma) \gtrsim \text{BR}(\tau \to e \gamma) \gtrsim  \text{BR}(\mu \to e \gamma)$ in both the NO and IO cases. 
We see that larger BRs tend to be predicted in the IO case as compared with the NO case, and 
only little parameter sets are allowed in the IO case. 
We note that if we adopt smaller $s_\chi$ such as $10^{-3}$, most of the parameter sets are excluded since the magnitude of $F^{AB}$ increases by one order resulting in larger BRs. 

\begin{figure}[t!]\begin{center}
\includegraphics[width=60mm]{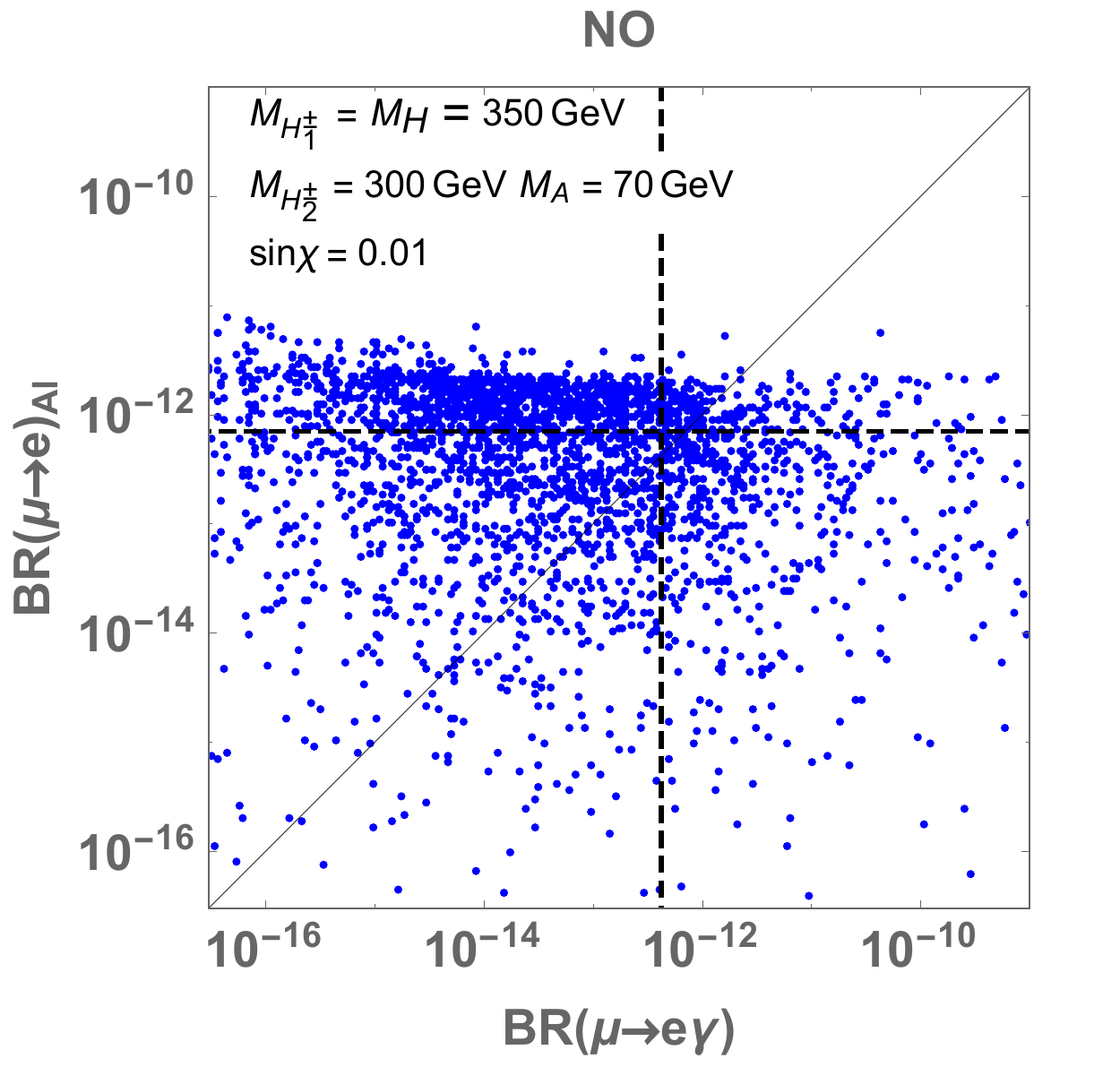} \qquad
\includegraphics[width=60mm]{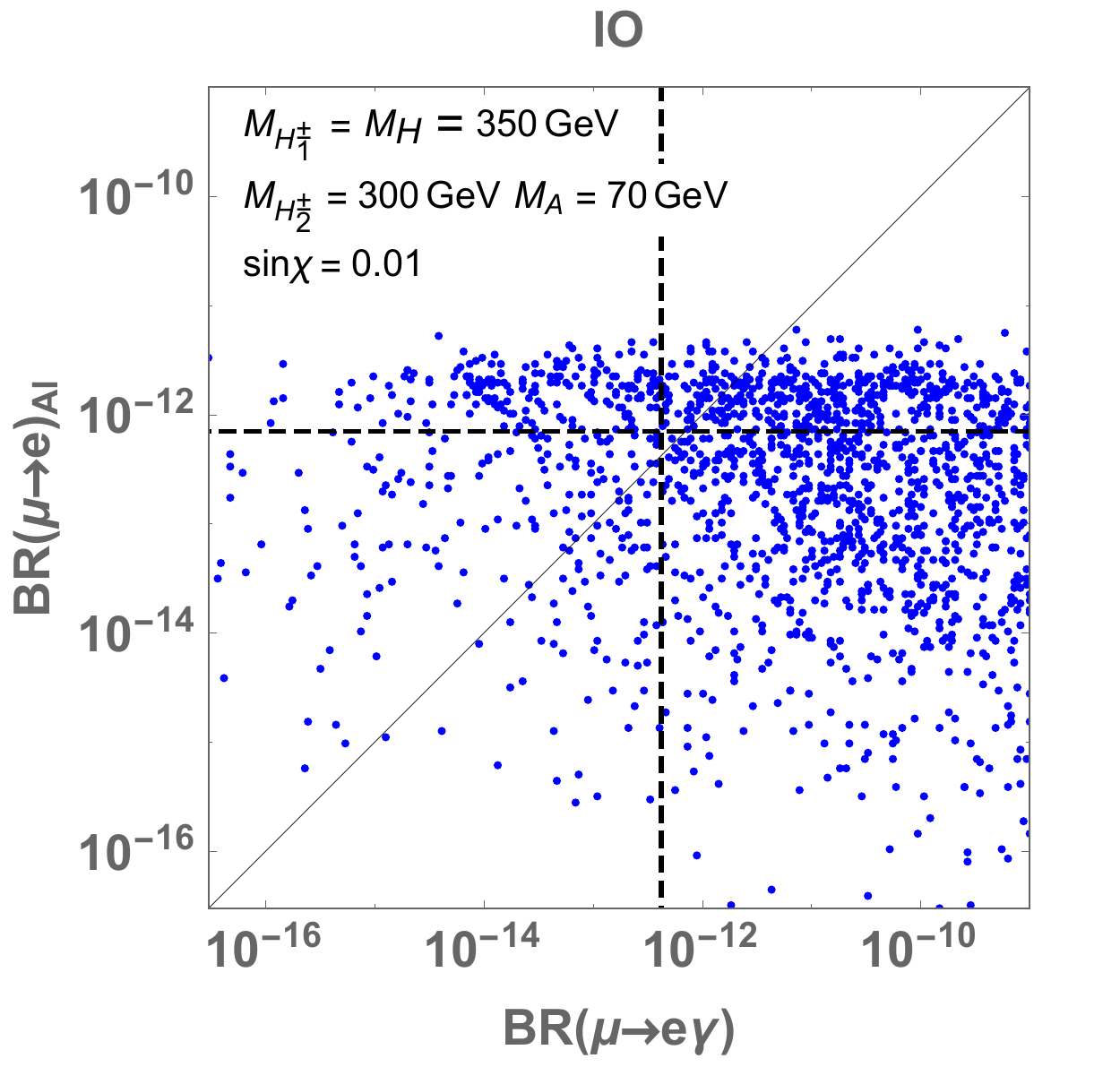} 
\caption{Correlations between BR$(\mu \to e \gamma)$ and BR$(\mu \to e)_{Al}$ for the case same as Fig.~\ref{fig:LFV1} where left(right) plots correspond to NO(IO).
The dashed horizontal and vertical lines indicate the upper bounds for each BR with 90\% confidence level. }   
\label{fig:MEC}
\end{center}
\end{figure}

In Fig.~\ref{fig:MEC}, we also show correlation between BR$(\mu \to e \gamma)$ and BR$(\mu \to e)_{Al}$ where the latter is $\mu$-e conversion rate for $^{27}_{13}$Al.
We find that $BR(\mu \to e)_{Al} >$ BR$(\mu \to e \gamma)$ in some region. 
These region correspond to the case where $(Y_{\Phi'})_{11} \ll (Y_{\Phi'})_{12(21)}$ so that $\mu \to e \gamma$ is more suppressed than $\mu$-$e$ conversion.

Before closing this section, let us briefly comment on the other observables in the lepton sector under the constraints from the CLFV data. 
In particular, we consider
the effective Majorana mass $\langle m_{ee} \rangle$ for the neutrinoless double beta decay $0 \nu \beta \beta$, which 
can be non-zero as follows: 
\begin{equation}
\langle m_{ee} \rangle = |m_1 \cos^2 {\theta_{12}} \cos^2 {\theta_{13}} + m_2 \sin^2 {\theta_{12}} \cos^2 {\theta_{13}} e^{i \varphi_1} + m_3 \sin^2 {\theta_{13}} e^{i(\varphi_{2} - 2 \delta_{\rm CPV})}|.
\end{equation} 
In our model, however, there is no particular prediction of the neutrino mixing angles, the squared mass differences and $\delta_{\rm CPV}$, so that 
$\langle m_{ee} \rangle$ can be obtained just by inputting the neutrino parameters which are suggested by global fit results. 
For example, 
taking the central values of the neutrino parameters and Eq.~(\ref{eq:inputs}), 
we obtain $\langle m_{ee} \rangle  < 0.05\,(0.07)$ eV and $\sum m_\nu \in [0.06, 0.18]\,([0.09, 0.20])$ eV for the NO (IO) case. 
This is allowed by the current bound from the KamLAND-Zen~\cite{KamLAND-Zen:2016pfg} experiment $\langle m_{ee} \rangle < \mathcal{O}(0.1)$ eV, while  
some points give the value of $\sum m_\nu$  larger than cosmological limit by PLANCK $\sum m_i < 0.12$ eV~\cite{Aghanim:2018eyx}. 
In future searches for the neutrinoless double beta decay, 
$\langle m_{ee} \rangle $ would be probed below about 0.05 eV~\cite{KamLAND-Zen:2016pfg}, and our model can be indirectly tested. 

\section{Collider Phenomenology of the Higgs Bosons\label{sec:collider}}

\begin{figure}[t!]\begin{center}
\includegraphics[width=75mm]{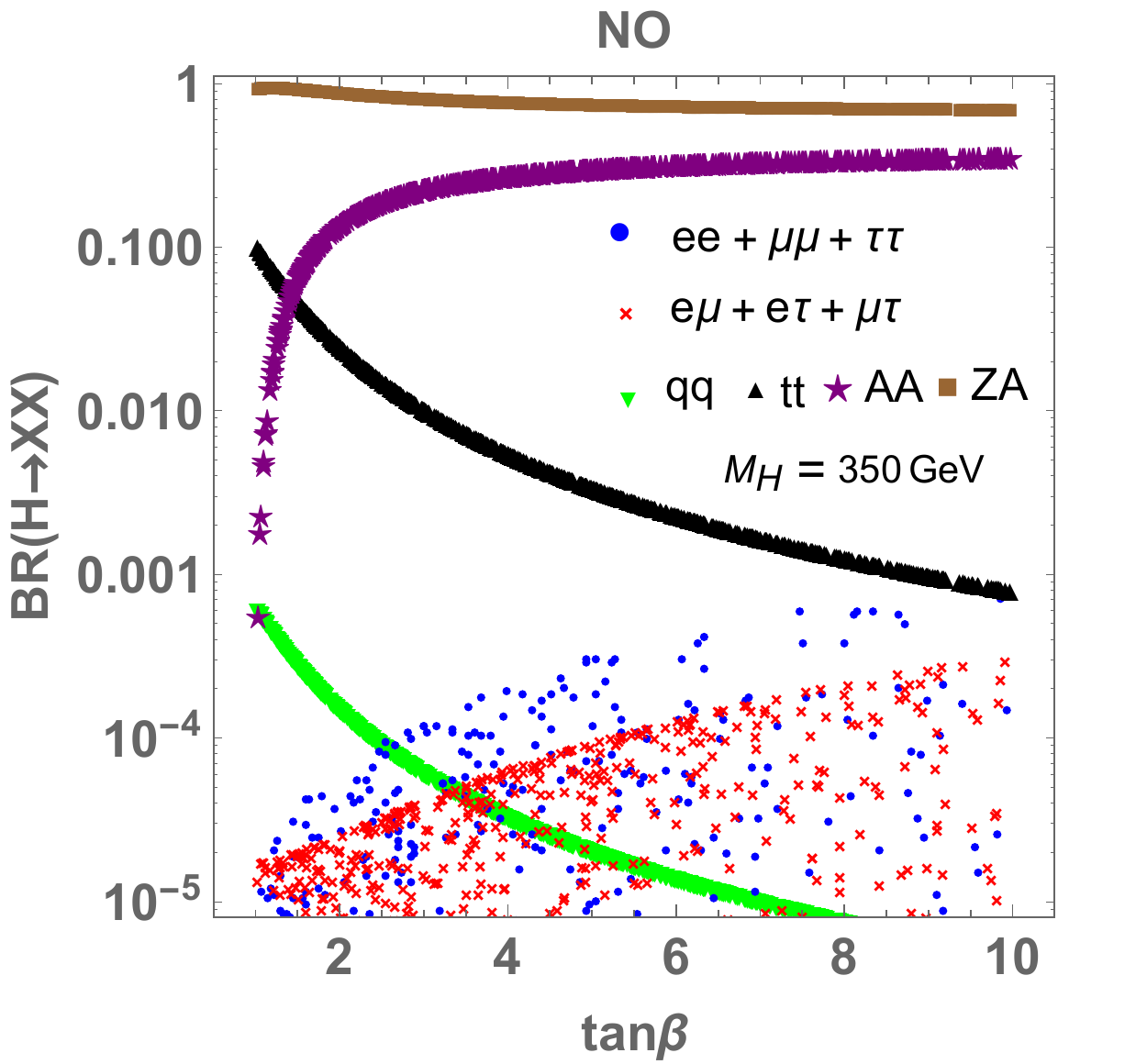} \quad
\includegraphics[width=75mm]{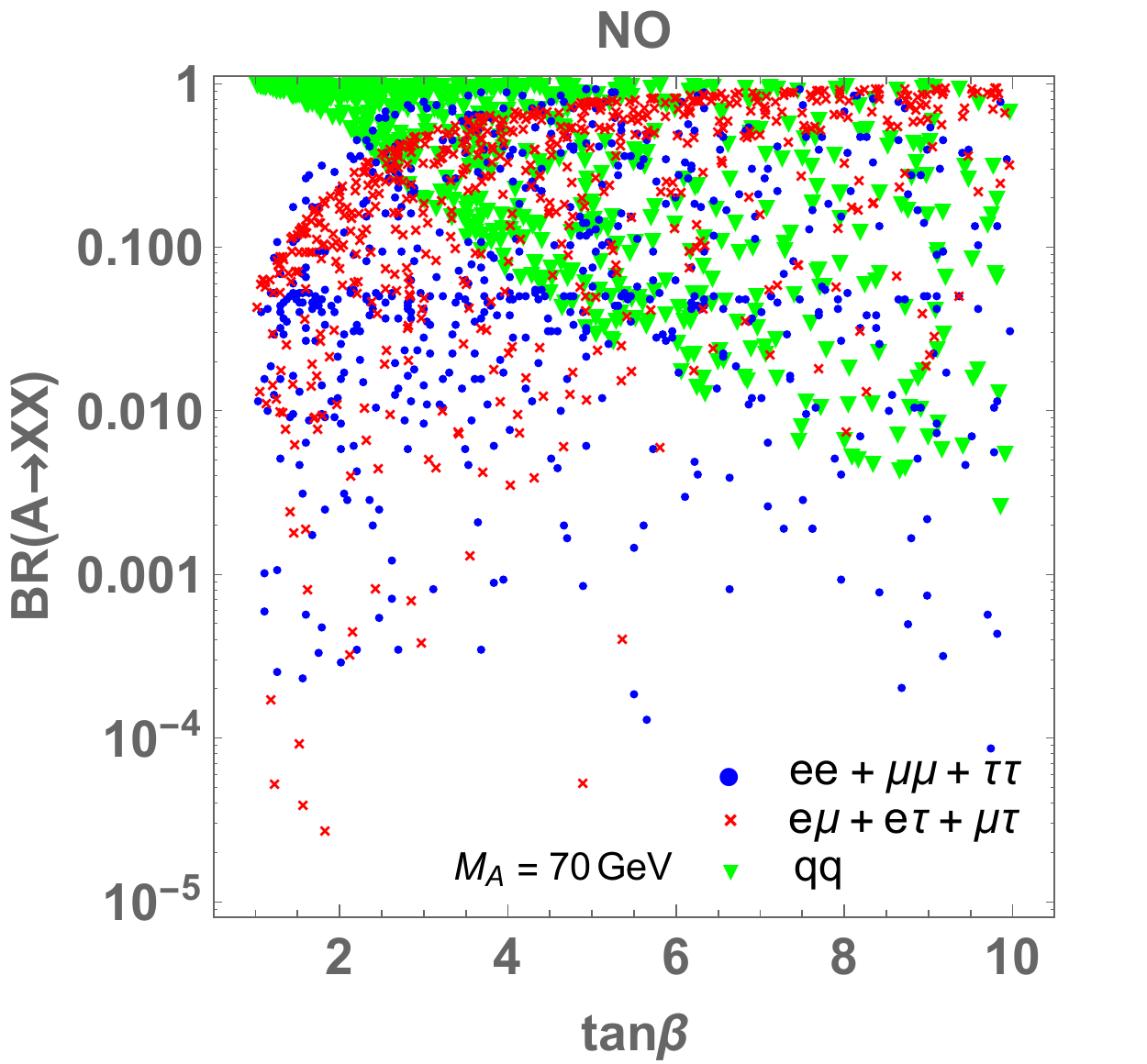}
\caption{BRs of $H$(left) and $A$(right) for the lepton flavor conserving modes ($ee+\mu\mu+\tau\tau$), the LFV modes ($e\mu+e\tau+\mu\tau$), the hadronic modes $(\sum_{q\neq t} q \bar q)$ modes, 
the $t\bar{t}$ mode and the $AA/ZA$ mode (only for $H$) as a function of $\tan \beta$; here $\ell \ell'$ indicate the sum of $\ell^+ \ell'^-$ and $\ell^- \ell'^+$ modes. 
The masses of $H$ and $A$ are fixed to be 350 GeV and 70 GeV, respectively.}   
\label{fig:LFV2}\end{center}\end{figure}

In this section, we discuss the collider phenomenology of our model, especially focusing on the lepton flavor violating (LFV) decays of the additional neutral Higgs bosons $H$ and $A$.
Such LFV decays are induced from the Yukawa matrices $Y_{\Phi}$ and $Y_{\Phi'}$, whose values are constrained by the charged lepton masses, neutrino oscillation data and the CLFV data, 
so that we expect the appearance of a characteristic pattern of the decay BRs. 
We here consider only the NO case adopting parameter sets satisfying CLFV constraints in previous section; IO case gives similar behavior of neutral scalar boson decay BRs.

As in the previous section, we scan the parameters written in Eq.~(\ref{eq:inputs}), and take 
$s_{\beta-\alpha} = 1$, $s_\chi = 0.01$, $\epsilon = 10^{-4}$ and the masses $(m_T^{},m_H^{},m_{H_1^\pm},m_{H_2^\pm},m_A^{})= (1000,350,350,300,70)$ GeV. 
Due to the choice of the mass spectrum, the $H \to ZA$ and $H \to AA$ decays can be important whose decay rates are calculated as 
\begin{align}
\Gamma(H \to ZA) & = \frac{g^2}{64 \pi \cos^2 {\theta_W}} \frac{m_H^3}{m_Z^2} 
\left[\left(1 - \frac{m_Z^2}{m_H^2} - \frac{m_A^2}{m_H^2} \right)^2 - \frac{4m_Z^2m_A^2}{m_H^4}\right]^{3/2},  \\
\Gamma(H \to AA) & = \frac{\lambda_{HAA}^2}{8\pi m_H}\sqrt{1 - \frac{4m_A^2}{m_H^2}},  
\end{align}
where $\theta_W$ is the Weinberg angle and $\lambda_{HAA} = (m_H^2 - m_A^2)\cot2\beta/v$. 
Differently from the leptonic decays of the Higgs bosons, 
decay rates into a quark pair are the same as those given in the Type-I THDM.
Thus, among the hadronic decays, the decay rates of $H/A \to b\bar{b}$ $(H/A \to t\bar{t})$ can be dominant for the Higgs boson mass below (above) twice of the top quark mass. 
We note that the decay modes of $H \to ZZ/W^+W^-/hh$ and $A \to Zh$ are absent in the alignment limit at tree level. 

\begin{figure}[t!]\begin{center}
\includegraphics[width=50mm]{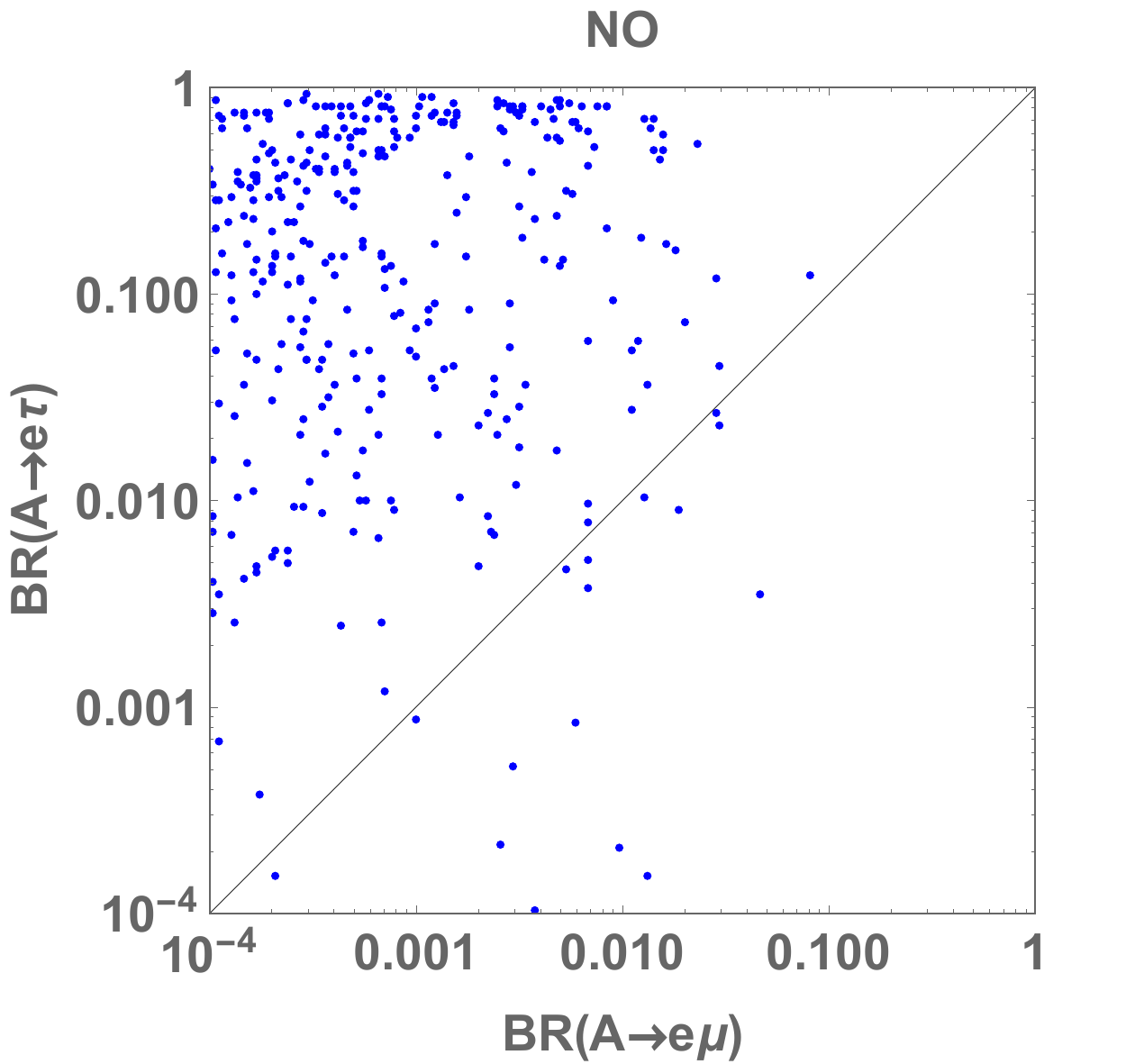}
\includegraphics[width=50mm]{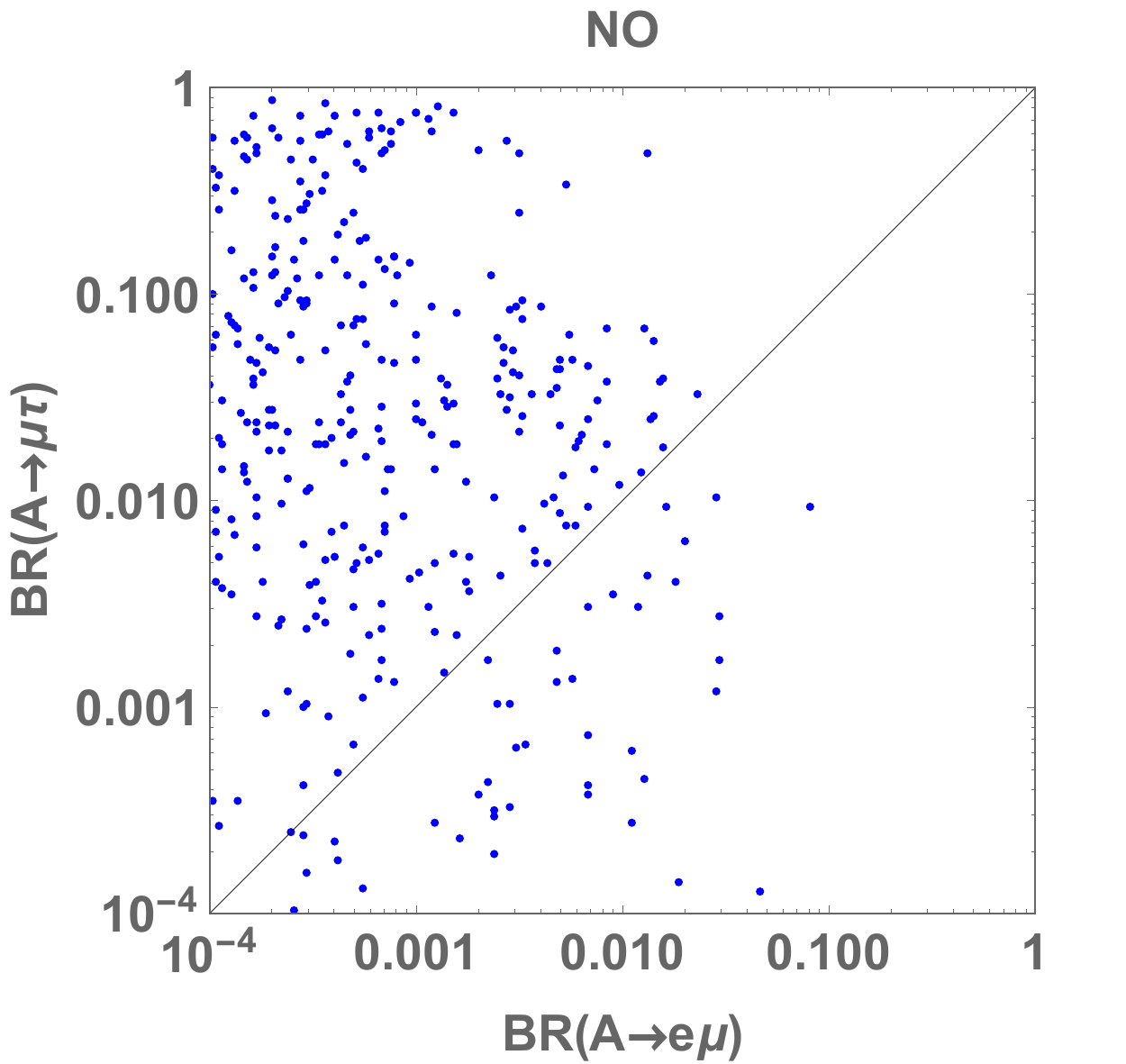}
\includegraphics[width=50mm]{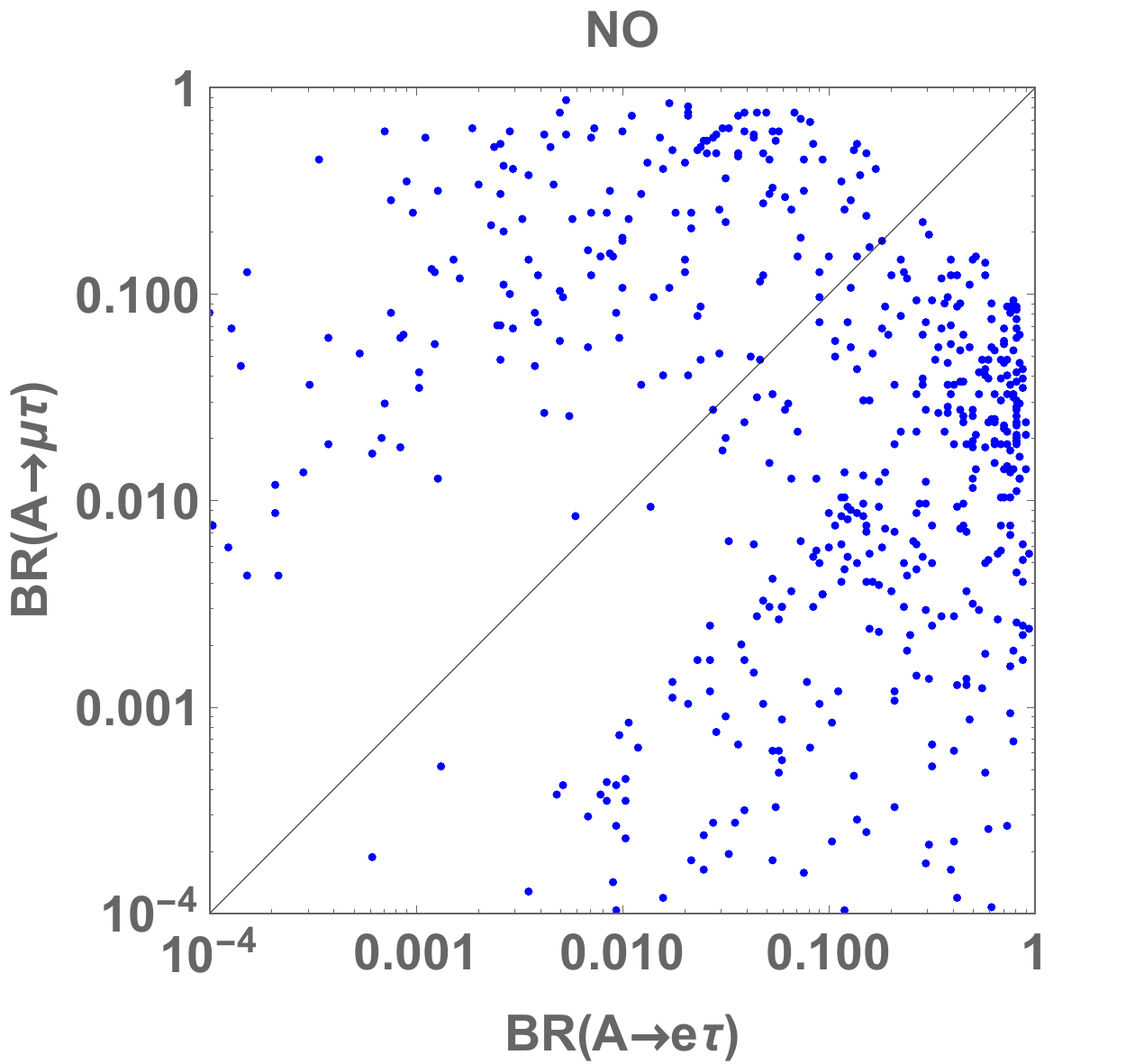}
\caption{Correlations among ${\rm BR}(A \to \ell_i \ell_j)$ shown where  
the mass of $A$ is fixed to be 70 GeV.}   
\label{fig:LFV3}\end{center}\end{figure}

In Fig.~\ref{fig:LFV2}, we show the BRs of $H$ and $A$ for the sum of the lepton flavor conserving modes, 
that of the LFV modes ($\ell_i \ell_j$ indicate the sum of $\ell^+_i \ell^-_j$ and $\ell^+_j \ell^-_i$), 
that of the hadronic modes (only the $t\bar{t}$ mode is separately shown) 
and the $AA/ZA$ modes (only for $H$) as a function of $\tan \beta$, fixing the masses to be $m_H = 350$ GeV and $m_A =70$ GeV.
The dominant mode of the $H$ decay is the $ZA$ mode, and the $AA$ or $t\bar{t}$ mode is the second largest one depending on $\tan \beta$, 
while the leptonic decay modes are smaller than $10^{-3}$.
The $H \to AA$ mode becomes significant for larger values of $\tan\beta$, because of the enhancement of the coupling $\lambda_{HAA} \sim  -(m_H^2 - m_A^2)\tan\beta/(2v)$ for $\tan\beta \gg 1$. 
For the $A$ decay, the LFV modes can be dominant for $\tan \beta \gtrsim 3$, since Higgs to Higgs decays and the $t\bar{t}$ mode are kinematically forbidden.
The correlations of the LFV decay modes of $A$ can be seen in Fig.~\ref{fig:LFV3}. 
We find that BR$(A \to e \mu)$ tends to be smaller than the other LFV decay modes, while both BR$(A \to e \tau) > $BR$(A \to \mu \tau)$ 
and BR$(A \to e \tau) < $BR$(A \to \mu \tau)$ cases can be found with similar amount.

We here comment on the decay of the SM-like Higgs boson $h$. 
In the THDMs, the couplings of $h$ coincide with those of the SM Higgs boson in the alignment limit at tree level. 
In our model, however, the Yukawa matrix $Y_\Phi$ is not exactly the same as the SM one even in the alignment limit, i.e., non-zero off-diagonal elements appear in the fermion mass eigenbasis. 
Such off-diagonal component is highly suppressed by the factor of $\epsilon^2$, so that the size of BRs of LFV decays of $h$ can be estimated by ${\cal O}(\epsilon^4)$. 
In fact, we numerically checked that the magnitude of BR$(h \to \ell\ell')$ is quite small, ${\cal O}(10^{-16})$ or less, in our benchmark scenario with $\epsilon = 10^{-4}$. 
This should be compared with the property of the additional Higgs bosons, in which the second Yukawa matrix $Y_{\Phi'}$ determines the decays of these Higgs bosons, and 
its off-diagonal elements are not suppressed by the $\epsilon$ factor, see Eq.~(\ref{eq:y0}).  

\begin{figure}[t!]\begin{center}
\includegraphics[width=70mm]{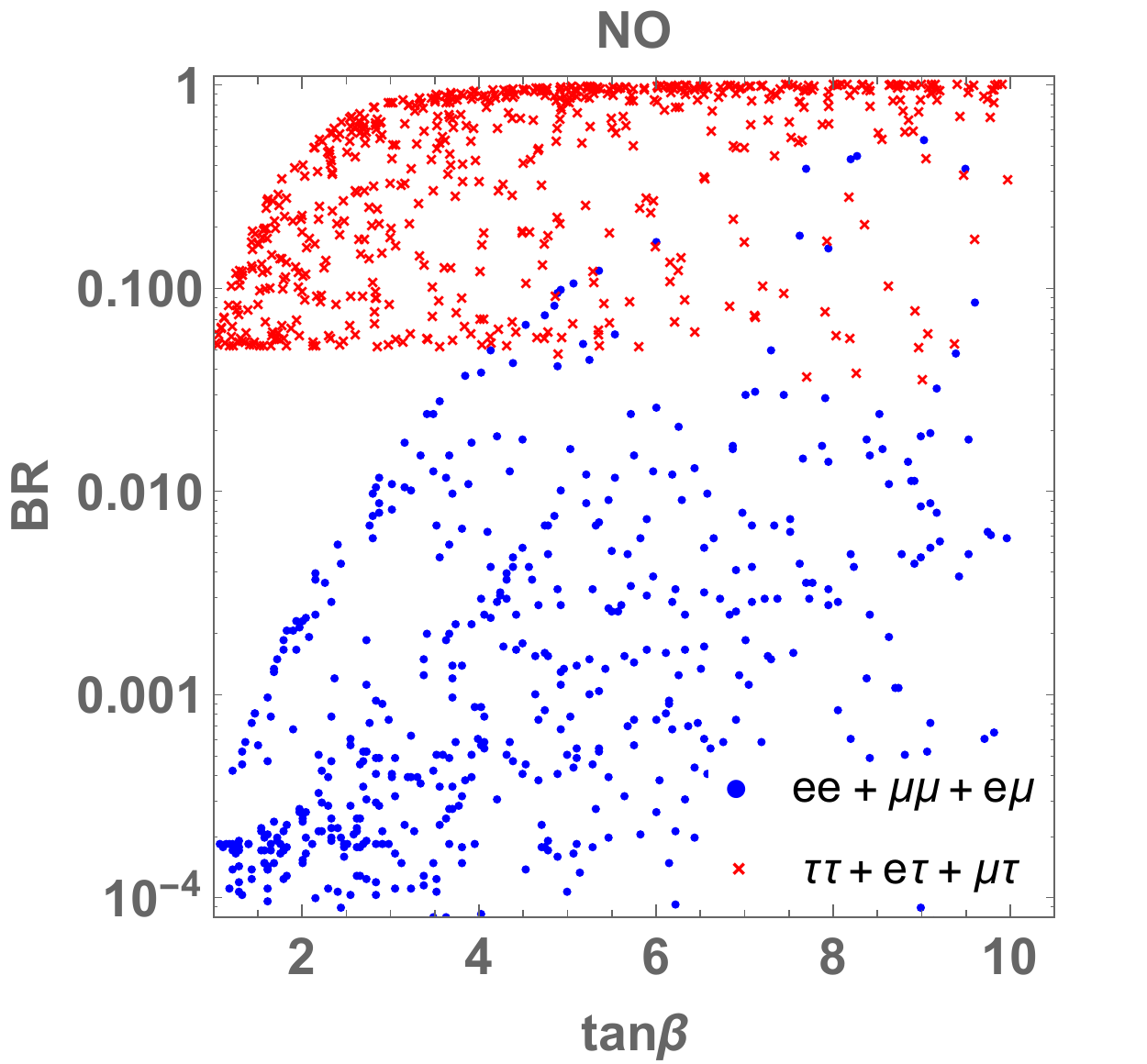}
\includegraphics[width=70mm]{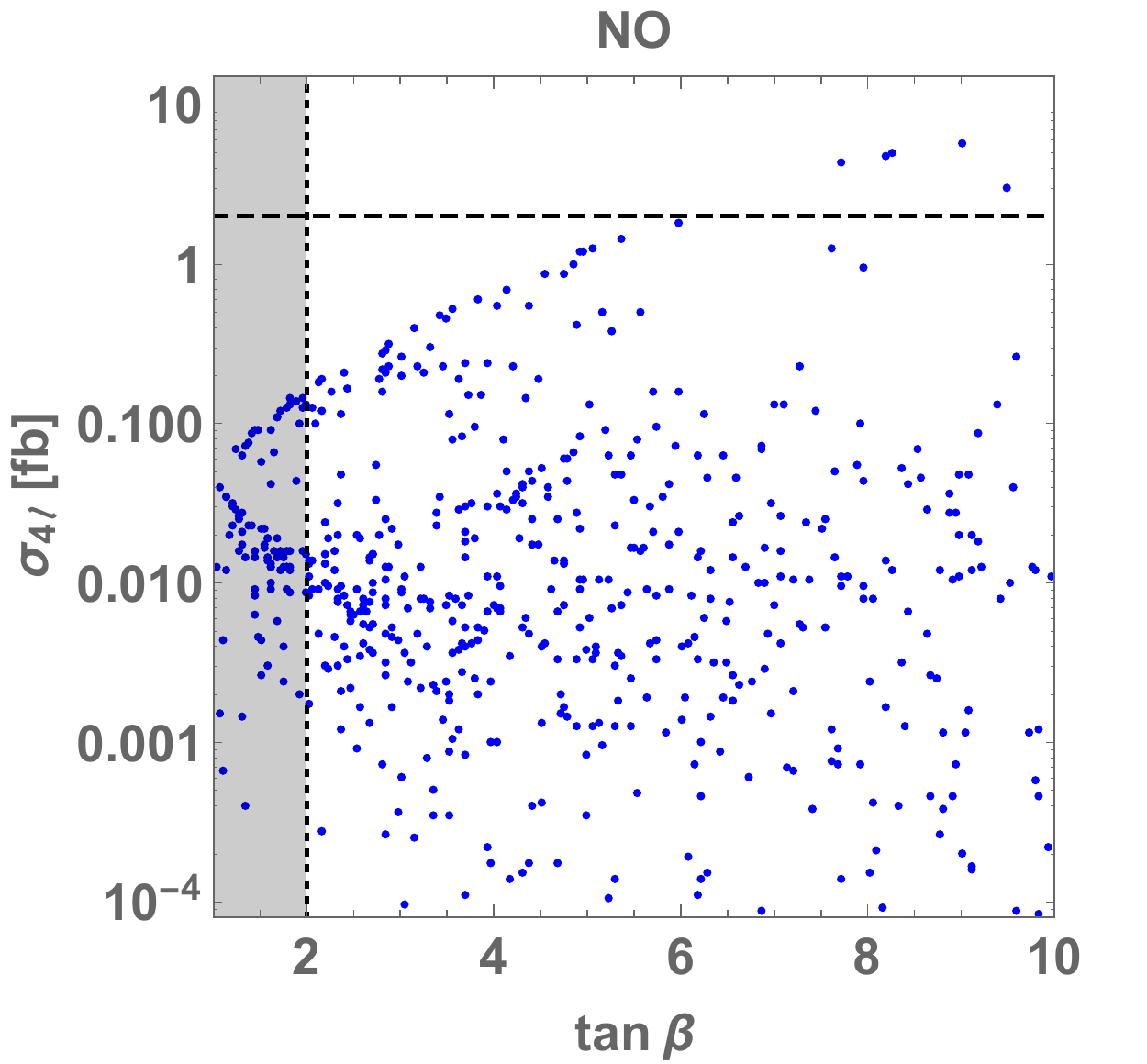}
\caption{(Left) Sum of the BRs of $A$ into $ee$, $e\mu$ and $\mu\mu$ (blue points) and that into $e\tau$, $\mu\tau$ and $\tau\tau$ (red points) as a function of $\tan\beta$ for $m_A = 70$ GeV. 
(Right) Cross section of the four lepton ($e$ or $\mu$) final states $\sigma_{4\ell}$ calculated by using Eq.~(\ref{eq:4l}) with $\ell =e$ or $\mu$ as a function of $\tan\beta$ for $m_A = 70$ GeV and $m_H=350$ GeV.
The dashed line indicates the current upper limit on the cross section given by the LHC data~\cite{Khachatryan:2017wny}.  
The shaded region is excluded by the constraint from $B_d \to \mu\mu$~\cite{Arbey:2017gmh,Haller:2018nnx}.
}
\label{fig:LFV4}\end{center}\end{figure}

Let us move on to the discussion of production processes of the additional Higgs bosons and constraints from current LHC data. 
At the LHC, the additional Higgs bosons can mainly be produced via the gluon fusion process $gg \to H/A$.\footnote{The associated process with bottom quarks 
$gg \to b\bar{b}H/b\bar{b}A$ cannot be important in our scenario, because the quark Yukawa couplings are suppressed by $\cot\beta$ as in the Type-I THDM. }
The pair productions $pp \to HA/H^\pm H/H^\pm A$ can also be used for smaller mass cases. 
So far, no discovery of significant signatures has been reported, and it has taken lower limits on their masses and upper limits on relevant coupling constants. 
See, e.g., Ref.~\cite{Aiko:2020ksl} for the recent analysis of the constraints from direct searches at the LHC in the THDM. 
We see in~\cite{Aiko:2020ksl} that in the alignment limit $s_{\beta-\alpha}=1$ with degenerate masses of the additional Higgs bosons, 
most of the parameter region has not been excluded yet, e.g., 
$\tan\beta \gtrsim 2$ is allowed in the Type-I THDM. 
It is important to mention here that these searches typically have sensitivities to relatively larger mass regions, e.g., above 100 GeV, so that 
they cannot be simply applied to cases for lighter additional Higgs bosons and/or the case with mass differences. 

From the above analysis, we see that $H$ can mainly decay into $ZA$ and $AA$, and $A$ further decays into a lepton pair. 
Thus, clear signals of $H$ can be obtained by the decay chain of $H \to ZA/AA \to 4\ell$, and
its cross section can be estimated by 
\begin{align}
\sigma_{4\ell} &\simeq \sigma(gg \to H) \times [\text{BR}(H \to ZA)\times \text{BR}(A \to \ell\ell)\times \text{BR}(Z \to \ell\ell) \notag\\
&+ \text{BR}(H \to AA)\times \text{BR}(A \to \ell\ell)^2], \label{eq:4l}
\end{align}
where $\sigma(gg \to H)$ is the gluon fusion cross section for $H$, which can be estimated to be $4~{\rm pb} \times \cot^2 \beta$ for $m_H = 350$ GeV~\cite{deFlorian:2016spz}.
A similar process with the same final states has been searched at the LHC, i.e., 
$pp \to Z' \to \phi \phi (\phi \to 2 \ell)$ process ($\ell = e, \mu$) with $Z'$ and $\phi$ being an extra neutral gauge boson and a neutral scalar boson, respectively. 
For $m_{Z'} = 350$ GeV which can be replaced by $m_H$ in our case, the upper limit has been set to be $\sigma_{4\ell} \sim 2$ fb~\cite{Khachatryan:2017wny}. 
In the left panel of Fig.~\ref{fig:LFV4}, we show the sum of the BRs of $A \to \ell\ell$ ($\ell=e$ or $\mu$) and that including at least one $\tau$.
We see that the former quantity rapidly increases as $10^{-3}$, $10^{-2}$ and $10^{-1}$ at $\tan\beta \simeq 2$, 4 and 6, respectively, so that 
at high $\tan\beta$ some parameter points can be excluded by the current LHC data. 
In fact, as seen in the right panel of Fig.~\ref{fig:LFV4}, 
a portion of points exceeds the current upper limit of $\sigma_{4\ell}$ at around $\tan\beta \simeq 8$, 
while the case with even larger $\tan\beta$ is not excluded because the cross section is suppressed by $\cot^2\beta$ and the maximal value of the BR is saturated. 
In addition excluded cases correspond to $(Y_{\Phi'})_{ij} \ (i[j]=1,2) \gtrsim (Y_{\Phi'})_{3a(a3)} \ (a=1,2,3)$ in which the BR of modes including $\tau$ are smaller than the modes including only $e$ and/or $\mu$.
Therefore, our model is allowed by the current LHC data except for quite a few cases, and could be test at future collider experiments. 
Moreover, LFV decay signals with $\tau$ would significantly improve the testability of the model, 
since the cross section $\sigma_{4\ell}$ including $\tau$ is much larger than the signals including only $e$ and/or $\mu$. 

Finally, let us give a comment on the phenomenology of the light CP-odd Higgs boson. 
For $m_A < m_h/2$, $h \to AA$ is allowed, and bounds on such BRs of the Higgs to Higgs decays have been studied in Ref.~\cite{Bernon:2014nxa} in the THDM. 
In our scenario with $m_A = 70$ GeV, these decays are kinematically forbidden. 
In the intermediate mass range, i.e., $62.5 \lesssim m_A \lesssim 100$ GeV, 
the search for additional Higgs bosons decaying into diphoton 
is available~\cite{CMS:2017yta} at the LHC, where the value of cross section times BR has been constrained to be smaller than about 0.1 pb with the 95\% confidence level. 
In our model, the gluon fusion production of $A$ with the mass of 70 GeV is given to be about 40 pb$\times \cot^2\beta$~\cite{deFlorian:2016spz}, while 
the BR of $A \to \gamma\gamma$ is given of order $10^{-4}$. 
Therefore, the cross section times BR is much below the current upper limit. 
At future collider experiments such as the High-Luminosity LHC~\cite{CMS:2013xfa,ATLAS:2013hta} and lepton colliders\footnote{At the LEP experiments, 
the light $A$ can be produced associated with a fermion pair, i.e., $e^+ e^- \to f\bar{f}A$. 
We have checked that its cross section is of order 10 (1) ab at $\sqrt{s}\simeq m_Z\, (200)$ GeV
for $m_A = 70$ GeV, $f = b$ and $\tan\beta \simeq 1$. Thus, almost no event is generated at LEP. }, e.g., the International Linear Collider (ILC)~\cite{Baer:2013cma,Asai:2017pwp,Fujii:2019zll}, 
the Circular Electron-Positron Collider (CEPC)~\cite{CEPC-SPPCStudyGroup:2015csa} and the Future Circular Collider (FCC-ee)~\cite{Gomez-Ceballos:2013zzn}, 
our scenario could be tested via the LFV decays of the Higgs bosons with the characteristic decay pattern. 
It goes without saying that dedicated studies are required to clarify the feasibility of such signatures, and
such analyses are beyond the scope of the present paper.

\section{Electroweak Phase Transition}
\label{sec:ewpt}

In this section, we consider the cosmological consequences of our scenario, particularly focusing on the electroweak phase transition. 
It has been known that in the electroweak baryogenesis scenario~\cite{Kuzmin:1985mm,Shaposhnikov:1987tw},  
the strongly FOPT is required to realize sufficient departure from thermal equilibrium in order to maintain non-zero baryon asymmetry of the Universe. 
The criteria for the strongly FOPT can be expressed by $\varphi_c/T_c \gtrsim 1$ or $\varphi_n/T_n \gtrsim 1$, where $T_c$ $(T_n)$ 
is the critical temperature providing degenerate vacua (nucleation temperature of electroweak bubbles from the bounce solution)
and $\varphi_c$ $(\varphi_n)$ is the order parameter at $T_c\, (T_n)$. 
Although we should use $\varphi_n/T_n$ rather than $\varphi_c/T_c$ for the estimation of the strength of the FOPT, 
we show both of them for comparison. 
Typically, these two valuables are almost the same with each other unless considering a mechanism of the supercooling scenario in which the thermal barrier does not disappear until much lower temperature $T_n \ll T_c$~\cite{Iso:2017uuu,vonHarling:2017yew}. 

It has been known that additional bosons can enhance the FOPT~\cite{Dolan:1973qd}, because their loop effects provide a positive contribution to a cubic field term 
of the effective potential at finite temperature, which makes a potential barrier higher at around the critical temperature. 
Indeed, it has been clarified that additional Higgs bosons in the THDM can make the FOPT stronger as a consequence of the non-decoupling effect if their masses mainly come from the VEV~\cite{Kanemura:2004ch}. 
This can also be described in such a way that a new dimensionful parameter $M$ which is irrelevant to the VEV and appears in the mass of Higgs bosons
is taken to be smaller than the physical mass parameter of the additional Higgs bosons. 
In our model, such $M$ parameter directly corresponds to the mass of $A$, see Eq.~(\ref{eq:masq}), so that 
$m_A^{}$ has to be smaller than the other masses of the additional Higgs bosons in order to enhance the FOPT. 
In Appendix~\ref{sec:finiteT}, we present detailed analytic expressions for the effective potential at finite temperature. 
We note that the loop effect of the charged singlet fields $S^\pm$ can not be dominant in our model, because of the 
smaller degrees of freedom $(N = 2)$ as compared with those of the doublets $(N = 8)$\footnote{In Refs.~\cite{Kakizaki:2015wua,Hashino:2016rvx,Ahriche:2018rao},  it has been shown that 
a larger number of singlet scalar fields significantly enhance their loop effects on the effective potential, and makes the FOPT stronger. }. 
We numerically find that the FOPT can be maximal when the mass of singlet-like Higgs boson $m_{H_2^\pm}$ is taken to be around 300 GeV in our scenario. 

In the following, we numerically evaluate the strength of the FOPT, i.e., $\varphi_c/T_c$ or $\varphi_n/T_n$ by using  {\tt CosmoTransitions}~\cite{Wainwright:2011kj}. 
We have checked that the behavior of the FOPT in the case of the THDM is consistent with the previous work~\cite{Basler:2016obg}. 
Our model predictions can then be simply obtained by taking $M = m_A$ in the THDM and modifying the charged Higgs sector by that with $H_1^\pm$ and $H_2^\pm$. 
As mentioned above, we take a smaller value of $m_A^{}$ but larger than $m_h/2$, e.g., 70 GeV, to avoid the constraint from $h \to AA$ at the LHC, see previous section. 
In addition, we take into account the constraints from perturbative unitarity and vacuum stability discussed in Sec.~\ref{sec:model} and 
flavor experiments~\cite{Arbey:2017gmh,Haller:2018nnx} such as $B_d \to \mu\mu$. 

\begin{figure}[!tbp]
  \centering
  \includegraphics[width=.5\hsize]{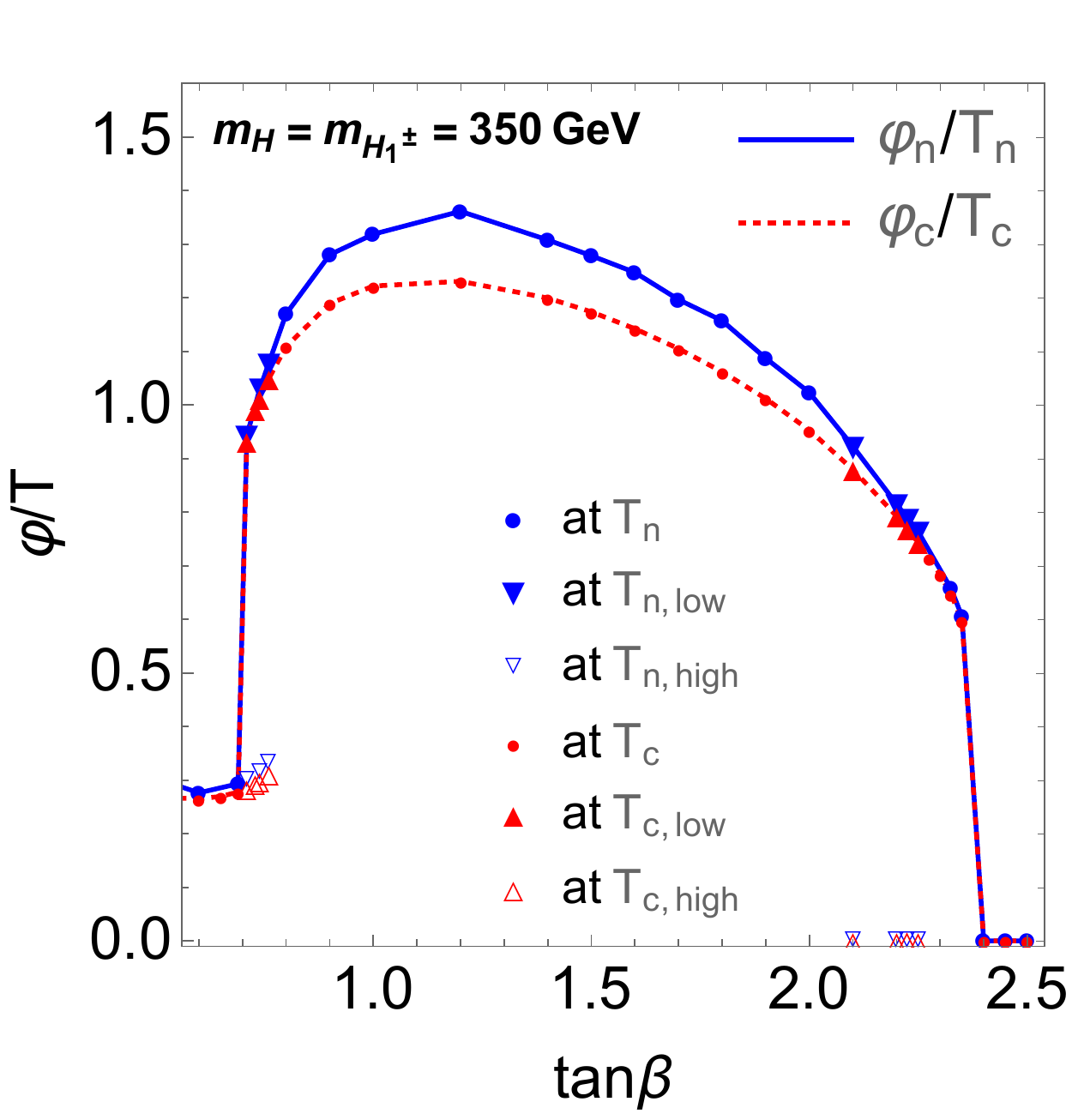}%
  \hfill
  \includegraphics[width=.5\hsize]{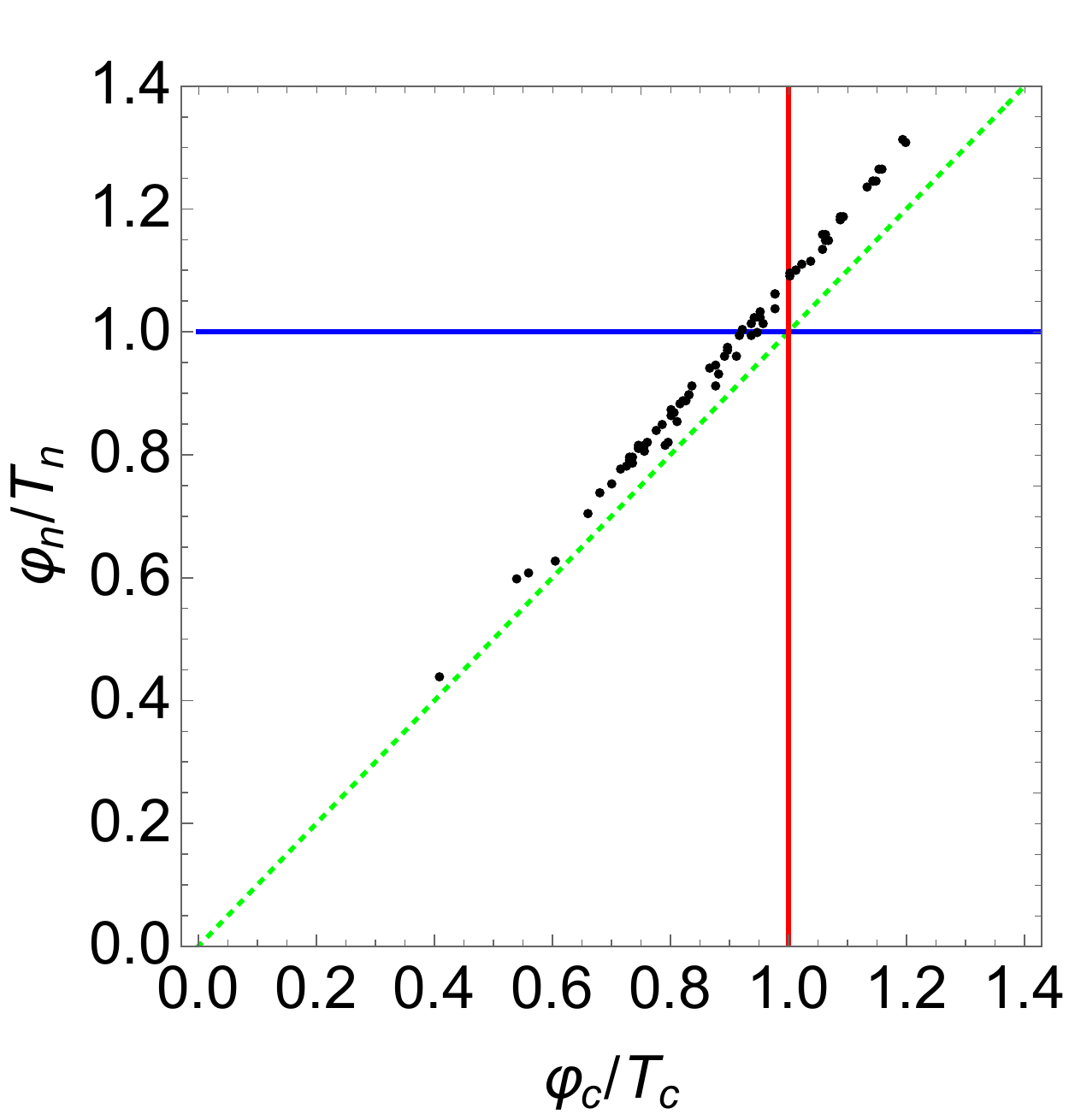}
  \caption{(Left) 
    Strength of the FOPT $\varphi/T$ at the critical temperature $T_c$ (red) and the bubble nucleation temperature $T_n$ (blue) as a function of $\tan\beta$ in the case 
    of $(m_A,m_H,m_{H_1^\pm},m_{H_2^\pm}) = (70,350,350,300)$ GeV, $m_S = 0$, $s_\chi = 0$, $\tan\sigma = 1$ and $s_{\beta-\alpha} = 1$. 
    At the circle (triangle) points, the one-step (two-step) phase transition is realized, 
    in which the empty (filled) triangles show the temperature at the first (second) phase transition denoted by $T_{c(n),\text{high}}$ ($T_{c(n),\text{low}}$) with $T_{c(n),\text{high}} > T_{c(n),\text{low}}$. 
    Points at $\varphi/T=0$ indicate the second-order phase transition. 
    (Right) Correlation between $\varphi_c/T_c$ and $\varphi_n/T_n$ with the scan of $m_H$ $(=m_{H_1^\pm})$ $\in$ [180, 500] GeV and $\tan\beta \in [1.4, 3.0]$. 
    All the other parameters are taken in the same way as those of the left plot. }
\label{fig:pt_tb}
\end{figure}

In Fig.~\ref{fig:pt_tb} (left), we show $\varphi_c/T_c$ and $\varphi_n/T_n$ as a function of $\tan\beta$ for the case with $m_A = 70$ GeV, $m_H=m_{H_1^\pm} = 350$ GeV and $m_{H_2^\pm} = 300$ GeV. 
We here neglect the effect of the small mixing angle between two charged Higgs bosons. For concreteness, we take $s_\chi = 0$.  
In this plot, we do not impose the constraints from the flavor experiments and the unitarity bound. 
We find that in some parameter points indicated by the triangle a two-step phase transition happens, where the empty (filled) triangles represent the value of 
$\varphi_c/T_c$ or $\varphi_n/T_n$ at the first (second) transition, while the circle points represent the case with the one-step phase transition.  
We see that both $\varphi_c/T_c$ or $\varphi_n/T_n$ become maximal at around $\tan\beta = 1.2$, and 
$\varphi_c/T_c$  $(\varphi_n/T_n)$ is greater than unity at $0.7 \lesssim \tan\beta \lesssim 1.9$ ($0.7 \lesssim \tan\beta \lesssim 2.0$). 
In addition, the value of $\varphi_n/T_n$ is slightly larger than $\varphi_c/T_c$, 
which can be more clearly seen from the right panel of Fig.~\ref{fig:pt_tb}, in which we show the correlation between $\varphi_c/T_c$ and $\varphi_n/T_n$. 
We see that  $\varphi_n/T_n$ is typically about 10\% larger than $\varphi_c/T_c$. 
This can mainly be explained by the difference between $\varphi_c$ and $\varphi_n$ ($\varphi_n > \varphi_c$) whose values are quite sensitive 
to the temperature at around the phase transition. 
In fact, around the $10\%$ difference between $\varphi_c$ and $\varphi_n$ arises from the slight difference between the temperatures, i.e.,  $(T_c-T_n)/T_c=\mathcal{O}(10^{-3})$. 

\begin{figure}[!tbp]
  \centering
  \includegraphics[width=.5\hsize]{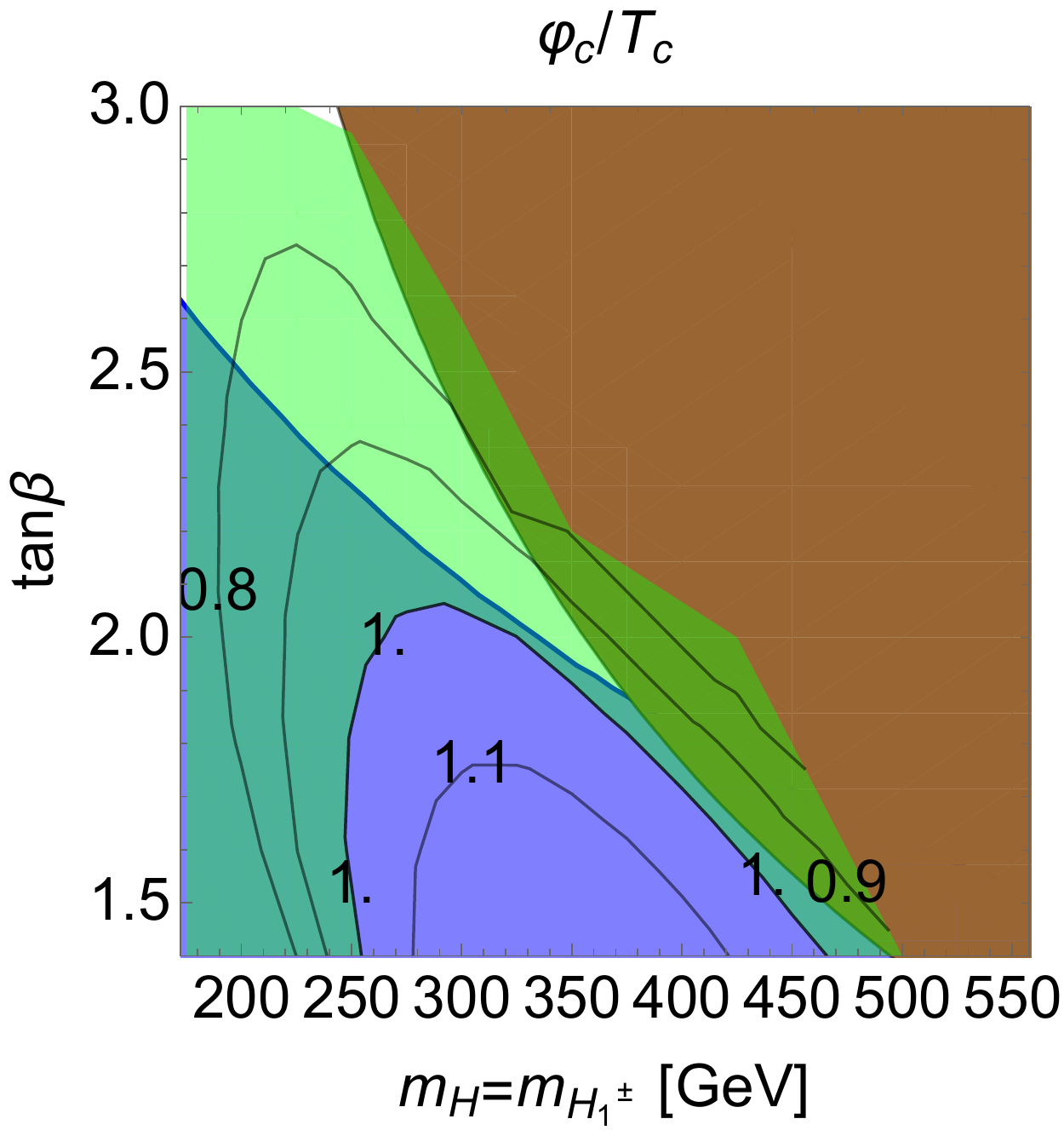}%
  \hfill
  \includegraphics[width=.5\hsize]{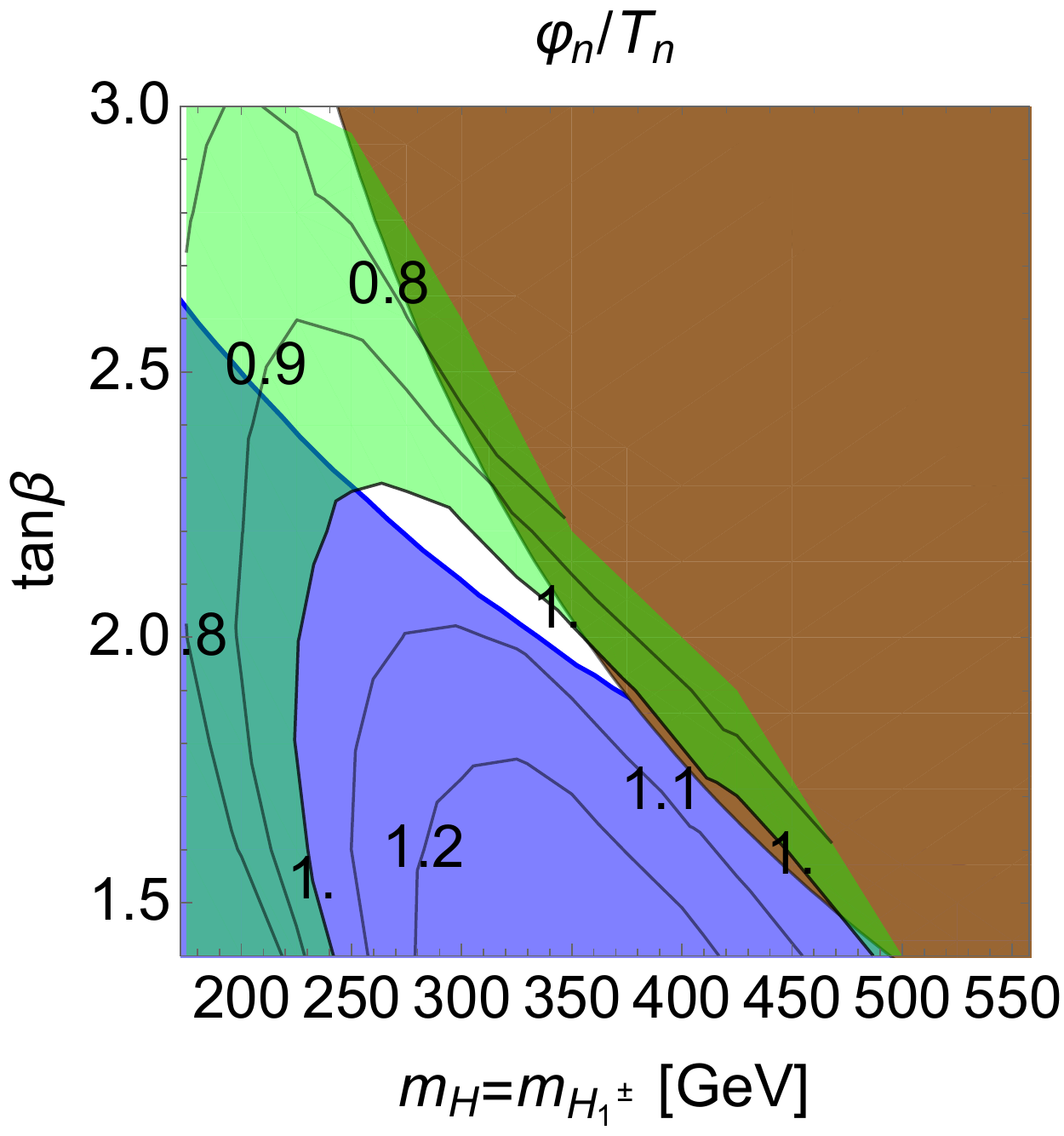}
  \caption{
    Contour plots of $\varphi/T$ at $T_c$ (left) and $T_n$ (right) in the $m_H (=m_{H_1^\pm})$-$\tan\beta$ plane, respectively.  
    We take a parameter set used in the right panel of Fig.~\ref{fig:pt_tb}.
    The green shaded region does not satisfy the criteria for the FOPT, i.e., $\varphi/T \gtrsim 1$.
    The blue and brown shaded region are respectively excluded by the constraint from $B_d \to \mu\mu$~\cite{Arbey:2017gmh,Haller:2018nnx} 
    and the unitarity bound given in Eq.~(\ref{eq:per-lam268}).
    }
  \label{fig:pt}
\end{figure}

In Fig.~\ref{fig:pt}, we show the contour plots for $\varphi/T$ at $T = T_c$ (left) and $T = T_n$ (right) in the $m_{H}^{} (= m_{H_1^\pm})$-$\tan\beta$ plane
using the same parameter set taken in the right panel of Fig.~\ref{fig:pt_tb}. 
Under the constraints from the flavor experiments and the unitarity bound, 
we find that there is a region of the parameter space which realizes the strongly FOPT ($\varphi_n/T_n \gtrsim 1$) shown as the white area in the right panel.  
It is clear that such region requires the quite limited parameter choice, i.e., $1.9 \lesssim \tan\beta \lesssim 2.3$ and $250 \lesssim m_H \lesssim 400$ GeV, which 
provides a good benchmark scenario for the collider phenomenology, as we discussed it in the previous section. 

It is worth discussing detectability of the gravitational waves (GWs) originated from the first-order phase transition since the typical frequency of GWs from the first-order phase transition at the electroweak scale is expected to explore by future planned space-based interferometers such as LISA~\cite{Seoane:2013qna}, DECIGO~\cite{Kawamura:2011zz} and BBO~\cite{Corbin:2005ny} which survey GWs in the millihertz to decihertz range. 
The GW spectrum can be parameterized by two dimensionless parameters, related to the released energy density $\epsilon$ and the change rate of the three dimensional bounce action $S_3$, defined as
\begin{align}
  \alpha \equiv \left. \frac{\epsilon}{\rho_\text{rad}} \right|_{T=T_n}^{}, \quad
  \widetilde{\beta} \equiv \left. \frac{d (S_3/T)}{d \ln T} \right|_{T=T_n},
  \label{eq:ab}
\end{align}
where $\rho_\text{rad}$ is the radiation energy density. 
We employ the analytic formula provided in Ref.~\cite{Caprini:2015zlo} to estimate the spectrum of the GWs, $\Omega_\text{GW} h^2(f)$. 
Parameters have one-to-one correspondence $(\tan\beta, m_H) \leftrightarrow (\alpha, \widetilde{\beta}) \leftrightarrow (f, \Omega_\text{GW} h^2)$ and correlations with the phase transition strength $\varphi_n/T_n \propto \Omega_\text{GW} h^2 \propto \alpha \propto \widetilde{\beta}^{-1}$.

\begin{figure}[!tbp]
  \centering
  \includegraphics[width=.5\hsize]{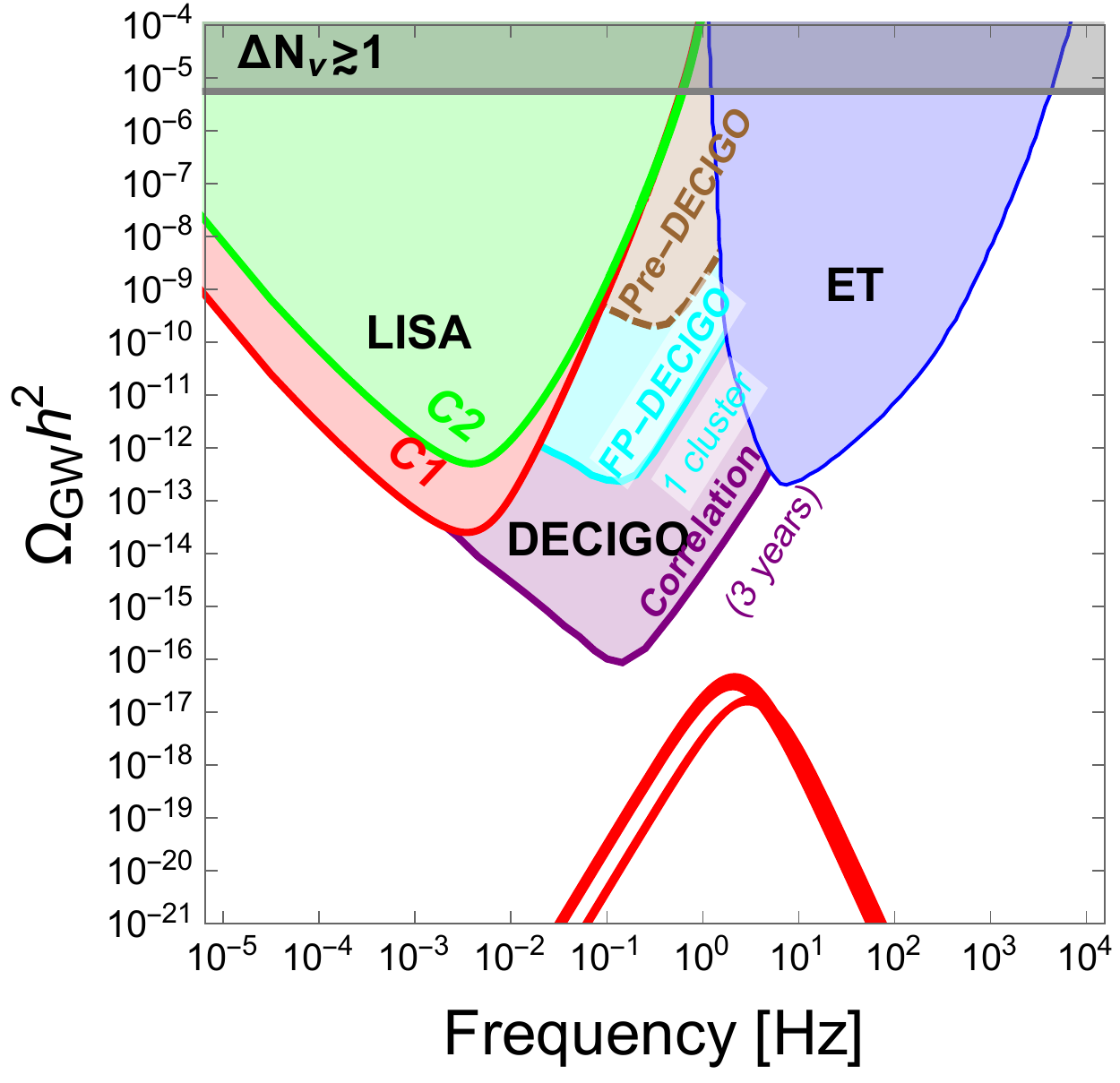}%
  \hfill
  \includegraphics[width=.5\hsize]{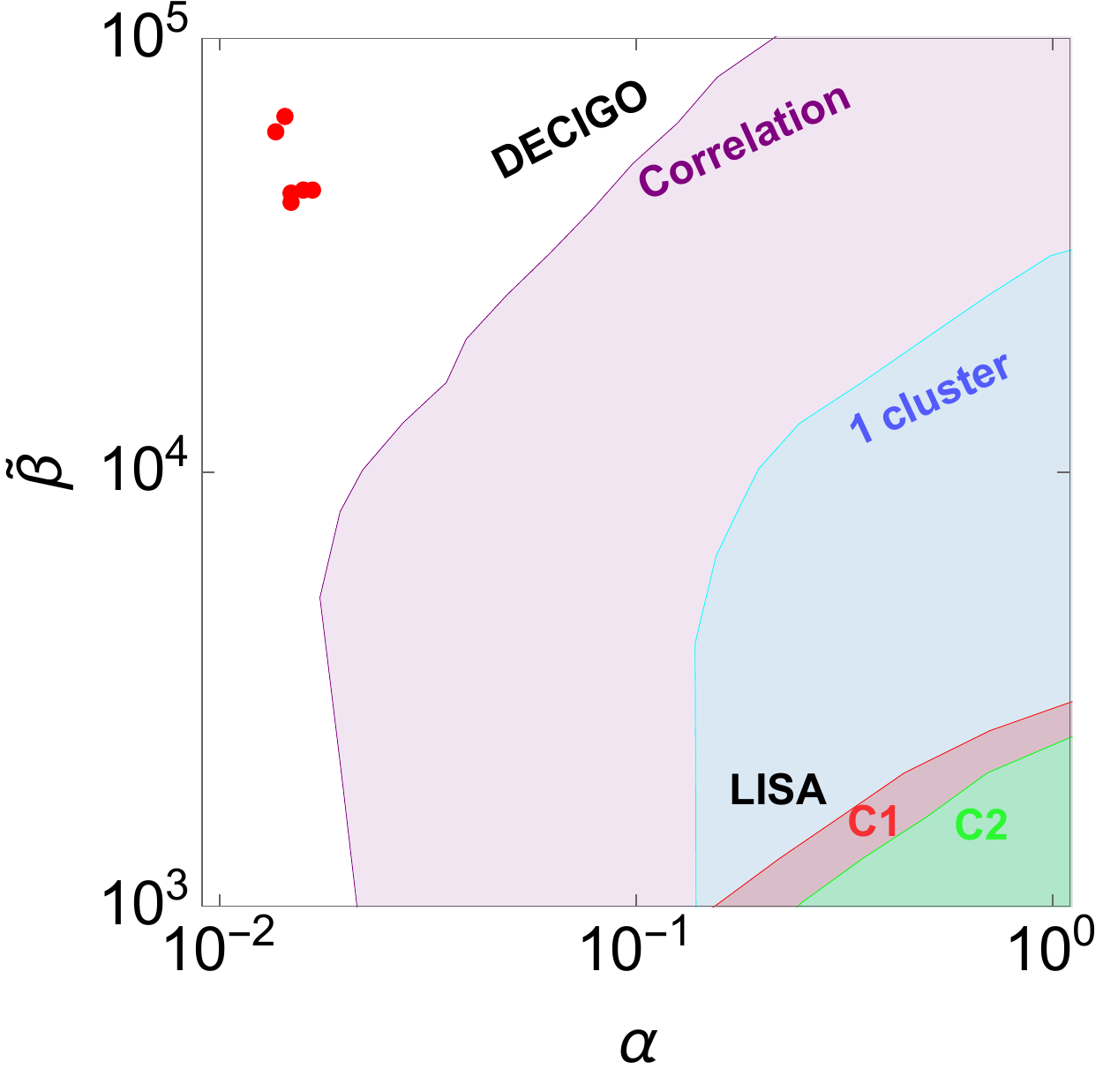}
  \caption{
    (Left) Predicted GW signals originated from the first-order phase transition are plotted with red lines, taking parameters on Fig.~\ref{fig:pt}. 
    The analytic formula provided in Ref.~\cite{Caprini:2015zlo} is used to estimate the spectrum of the GWs assuming the bubble wall velocity $v_b$ = 0.95 as an example. 
    Sensitivity curves for future GW interferometers, LISA~\cite{Caprini:2015zlo}, DECIGO~\cite{Kawamura:2011zz}, and Einstein telescope (ET)~\cite{Schmitz:2020syl} are plotted, respectively.
    (Right) Predicted values of $\alpha$ and $\widetilde{\beta}$ which derive the plot on the left panel. 
    Experimental sensitivities are obtained for the sound wave contribution fixing $T_n$ = 100 GeV. 
  }
  \label{fig:gw}
\end{figure}

In Fig.~\ref{fig:gw}, the predicted GW signals are plotted taking parameters on Fig.~\ref{fig:pt}. 
Red colors on each panel are allowed by the constraint from $B_d \to \mu\mu$. 
We find that the GW signals do not reach the future planned sensitivities of observations. 
Because the exact non-decoupling limit does not allow due to the global U(1) and the constraint from $B_d \to \mu\mu$ excluds most of the parameter space, the GW signals correlated with the strength of the first-order phase transition cannot be enough strong.  

\begin{figure}[t!]\begin{center}
\includegraphics[width=70mm]{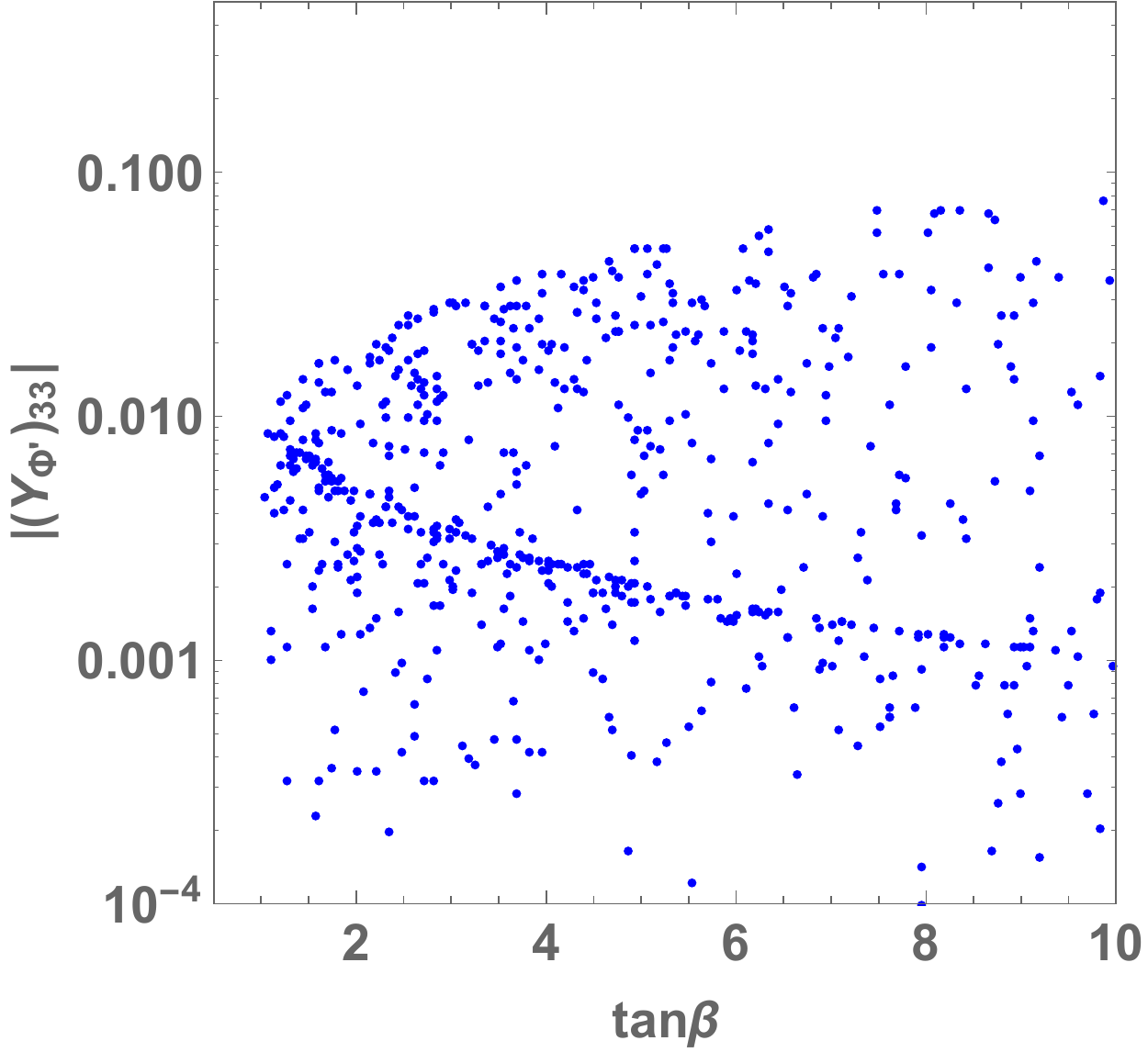}
\includegraphics[width=70mm]{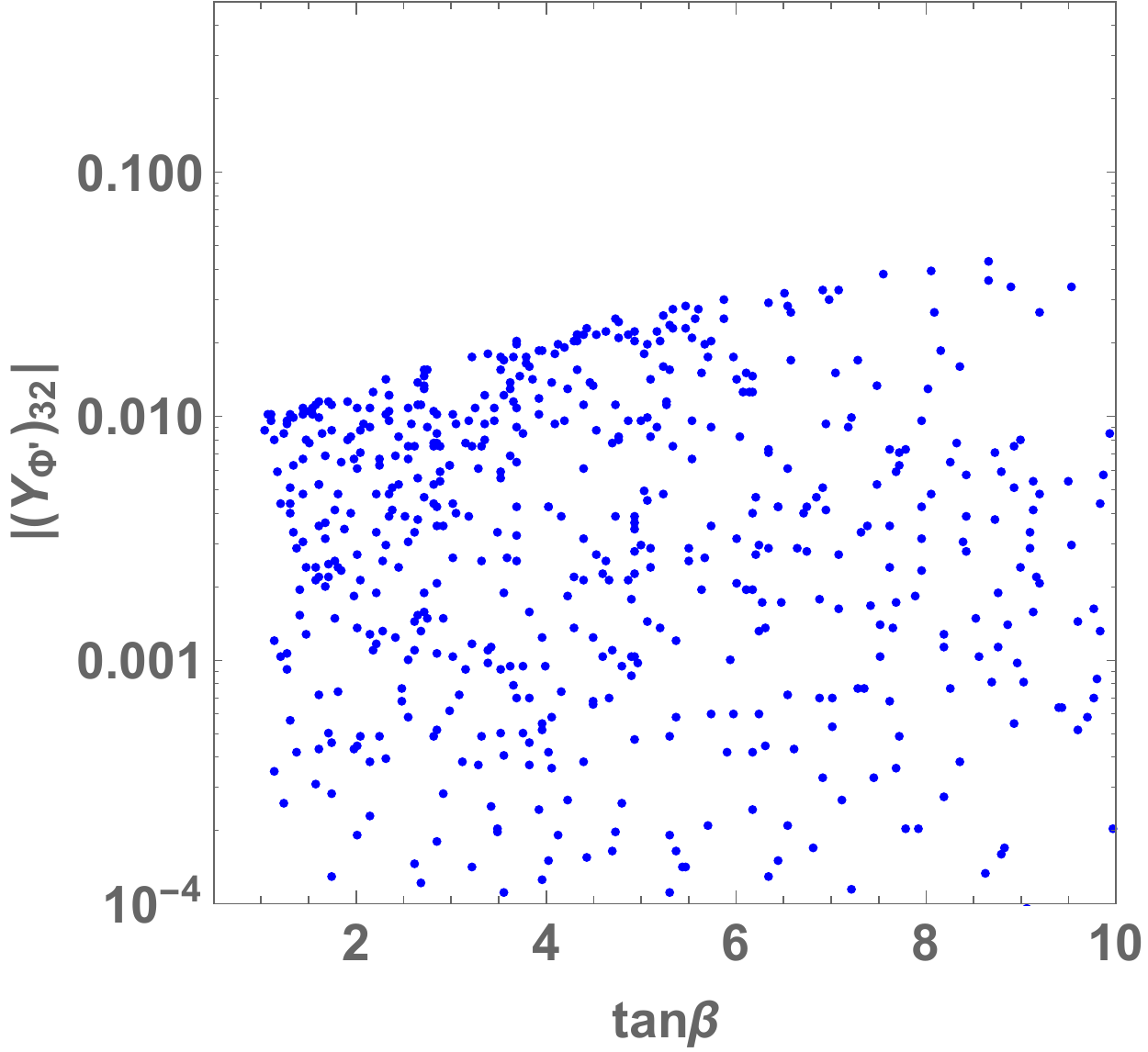}
\caption{Values of $(Y_{\Phi'})_{33}$ (left) and $(Y_{\Phi'})_{32}$ (right) couplings as a function of $\tan\beta$.}
\label{fig:ewbg}
\end{center}
\end{figure}

Finally, we give a comment on the possibility for the generation of baryon asymmetry of the Universe. 
As it is known that the strongly FOPT is one of the necessary conditions to obtain the non-zero baryon number in the electroweak phase transition. 
In order to estimate the generated baryon number, we also need to calculate the reflection rate of chiral fermions against the bubble, 
in which non-zero charges, e.g., the hypercharge and isospin, can be accumulated at the vicinity of the bubble if CP-violation happens in the interaction among the bubble and fermions.  
Such non-zero charges can be converted into the baryon number from the sphaleron process, which is frozen in the broken phase because of the decoupling of the sphaleron process. 
See e.g.,~\cite{Funakubo:1996dw} for more detailed discussions. 
In Ref.~\cite{Chiang:2016vgf}, the electroweak baryogenesis has been discussed in a two Higgs doublet model by introducing CP-violating phases in lepton Yukawa couplings, particularly 
$\tau$-$\tau$ and $\tau$-$\mu$ couplings to additional Higgs bosons. 
It has been shown that in order to explain the observed baryon asymmetry the magnitude of these Yukawa couplings has to be of order 0.1--1. 
We thus plot the values of corresponding Yukawa couplings $(Y_{\Phi'})_{33}$ and $(Y_{\Phi'})_{32}$ in our model in Fig.~\ref{fig:ewbg}. 
It is seen that in the region where the FOPT is realized, i.e., $\tan\beta \simeq 2$, the value of $(Y_{\Phi'})_{33}$ and $(Y_{\Phi'})_{32}$ can be 
maximally of order 0.01. 
Therefore, qualitatively our model can generate baryon asymmetry which is one or two orders of magnitude smaller than the required value, so that 
additional sources of baryon asymmetry would be needed.


\section{Conclusions\label{sec:conclusion}}

We have investigated the Zee model which is extended by introducing a vector-like lepton doublet and a flavor dependent global $U(1)'$ symmetry. 
Because of the $U(1)'$ symmetry, FCNCs in the quark sector can naturally be avoided at tree level, while a sufficient new source of LFV interactions is provided, which 
is required to explain the current neutrino oscillation data. 
In particular, we have focused on the weak mixing scenario, where the fourth charged lepton is weakly mixed, with a magnitude of $\epsilon(\ll 1)$, with three generations of the SM leptons. 
This scenario can successfully suppress CLFV decays, but give an important new contribution to the neutrino mass matrix. 
By scanning model parameters under which neutrino oscillation data are satisfied, 
we have shown the appearance of strong correlations among the BRs of CLFV decays, i.e., 
$\text{BR}(\tau \to \mu \gamma) \gtrsim \text{BR}(\tau \to e \gamma) \gtrsim  \text{BR}(\mu \to e \gamma)$ in both the NO and IO cases. 

Our model provides additional sources of CP-violation in the lepton sector. 
We thus have studied an impact of these CP-violating phases on the electron EDM. 
It is shown that the new contribution to the electron EDM is sufficiently smaller than the current experimental limit, because of the 
structure of the Yukawa matrices constrained by the $U(1)'$ symmetry and the weak mixing scenario. 
Thus, our scenario is allowed even if we take ${\cal O}(1)$ phases of the lepton sector. 

We then have discussed the collider phenomenology, especially focusing on the LFV decays of the neutral Higgs bosons. 
In our benchmark scenario motivated by realization of the strongly FOPT, 
the additional CP-even Higgs boson $H$ mainly decays into the light CP-odd Higgs boson $A$ such as $ZA$ or $AA$, and 
$A$ dominantly decays into a lepton pair with both lepton flavor conserved and violated one. 
Thus, the four lepton final states are expected at the LHC via $pp \to H \to ZA/AA \to \ell_i\ell_j\ell_k\ell_l$. 
We have confirmed that the cross section of this process is below the current upper limit driven by the LHC data, except for 
a small portion of the parameters allowed by the CLFV data with $\tan\beta \simeq 8$. 
This signature can be a smoking gun to probe our model, and would be tested at future collider experiments such as the HL-LHC and future lepton colliders. 

Finally, as a cosmological consequence, we have studied the electroweak phase transition by using the effective potential at finite temperature. 
The content of the Higgs sector in our model is the same as that of the original Zee model, i.e., two Higgs doublets and a pair of charged singlet scalars. 
Due to the $U(1)'$ symmetry, the coefficient of the $(\Phi_1^\dagger \Phi_2)^2$, denoted as $\lambda_5$, is forbidden, and thus
the CP-odd Higgs boson $A$ becomes a pseudo-NG boson whose mass arises from the $(\Phi_1^\dagger \Phi_2)$ term.  
Because of this property, we need a light $A$ in order to make non-decoupling effects of the additional Higgs bosons stronger, which 
plays an important role to realize the strongly FOPT. 
We have found that the strongly FOPT can be realized in the case where 
$1.9 \lesssim \tan\beta \lesssim 2.3$ and $250 \lesssim m_H \lesssim 400$ GeV for $m_A = 70$ GeV, which 
is compatible with the constraints from the flavor experiments and LHC data. 
Therefore, we have shown that our model contains two important ingredients to realize the successful electroweak baryogenesis scenario, i.e., 
the additional CP-violating phases and the strongly FOPT. 
According to the qualitative discussion, our model can generate baryon asymmetry which is one or two orders of magnitude smaller than its observed amount, so that 
additional sources of baryon asymmetry would be required. 

\begin{acknowledgments}

The authors would like to thank Dr. Eibun Senaha for fruitful discussions about the electroweak baryogenesis 
and Dr. Hiroshi Okada for giving us an important comment on neutrino masses. 
The work of KY is supported in part by the Grant-in-Aid for Early-Career Scientists, No.~19K14714. 
TM is supported by National Research Foundation of Korea under Grant Number 2018R1A2B6007000. 

\end{acknowledgments}

\appendix

\section{Approximate Formulae for the $F$ Matrix \label{sec:fmat}}

In Sec.~\ref{sec:neutrino}, we show that the matrix elements $F$ can be expressed in terms of the neutrino mass matrix by solving Eq.~(\ref{eq:dat}).  
Neglecting the electron mass except for terms proportional to $1/m_e$, we find the following approximate formulae for the elements of 
the $F$ matrix: 
\begin{align}
\begin{split}
F_{12} &\simeq  \frac{m_{\rm dat}^{11}}{2  C_{\rm diag}^0 m_\mu }\frac{ \delta_{24} y_{33}-y_{23} \delta_{34}}{y_{12}\delta_{24} 
   y_{33} +\delta_{14} y_{23} y_{32} +  y_{13} y_{22}\delta_{34}- y_{12} y_{23}\delta_{34} - y_{13}\delta_{24}
   y_{32}-\delta_{14} y_{22} y_{33}}, \\
F_{13} &\simeq -\frac{m_{\rm dat}^{11}}{2C_{\rm diag}^0 m_\tau}
\frac{ \delta_{24} y_{32} - \delta_{34} y_{22}}{y_{12} \delta_{24} y_{33}-  y_{12} y_{23}\delta_{34} + y_{13} y_{22}\delta_{34} - y_{13}\delta_{24}y_{32}-\delta_{14} y_{22} y_{33}+\delta_{14} y_{23} y_{32}}, \\
F_{23} &\simeq -\frac{m_{\rm dat}^{33} \delta_{24}^2  -2 m_{\rm dat}^{23}\delta_{24} \delta_{34} + m_{\rm dat}^{22}\delta_{34}^2 }
{2 C_{\rm diag}^0 \left[m_\mu\delta_{24}  (\delta_{24} y_{32}- y_{22}\delta_{34})+m_\tau\delta_{34}  (\delta_{24} y_{33}- y_{23}\delta_{34})\right]}, \\
F_{14} &\simeq \frac{m_{\rm dat}^{11}}{2 C_{\rm diag}^4 m_{T}}\frac{ y_{23} y_{32}-y_{22} y_{33}}{ y_{12}\delta_{24} 
   y_{33} +\delta_{14} y_{23} y_{32} +  y_{13} y_{22}\delta_{34}- y_{12} y_{23}\delta_{34} - y_{13}\delta_{24}
   y_{32}-\delta_{14} y_{22} y_{33}}, \\
F_{24} &\simeq \frac{m_{\rm dat}^{22}(\delta_{24}y_{32}r - y_{22}\delta_{34}r + \delta_{34}y_{33})  + y_{23}\delta_{24} m_{\rm dat}^{33}
-2m_{\rm dat}^{23}y_{23} \delta_{34}    }{2 C_{\rm diag}^4  m_{T} \left[( r \delta_{24} (\delta_{24} y_{32}-\delta_{34} y_{22})+ \delta_{34}(\delta_{24} y_{33}- y_{23}\delta_{34})\right]}, \\
F_{34} &\simeq \frac{m_{\rm dat}^{33}(\delta_{24}y_{33}-y_{23}\delta_{34}-  ry_{22}\delta_{24})+2r  m_{\rm dat}^{23} \delta_{24}y_{32}-rm_{\rm dat}^{22} y_{32}\delta_{34}}{2 C_{\rm diag}^4  m_{T} 
\left[r\delta_{24}    (\delta_{24} y_{32} -  y_{22}\delta_{34}) + \delta_{34}(\delta_{24} y_{33}-\delta_{34} y_{23})\right]}, 
\end{split}
\end{align}
where $y_{ij} \equiv (Y_{\Phi'}^0)_{ij}$, $\delta_{ij} \equiv (\delta Y_{\Phi'})_{ij}$ and $r \equiv m_\mu/m_\tau$. 
The matrix elements $m_{\rm dat}^{ij}$ denote those of the right-hand side of  Eq.~(\ref{eq:nudat}). 
These expressions agree with the exact ones typically within a 10\% error. 

\section{Decay Amplitudes for $\ell_i \to \ell_j \gamma$ Processes \label{sec:LFV_amp}}

Here, we summarize the analytic formulae for the amplitudes of the $\ell_i \to \ell_j \gamma$ processes denoted by $(a_{L,R})_{ij}$ in Eq.~(\ref{br_clfv}). 
The contributions from diagrams with the charged Higgs bosons inside loop are obtained as  
\begin{align}
(a_{R}^{H^\pm_1})_{ij} &= \frac{1}{16 \pi^2} \sum_{A=1}^4 \left[ (Y_{\Phi'})_{Aj}^{*} (Y_{\Phi'})_{Ai} c_\chi^2 F_1(m_{\ell_i}, m_{\ell_j}, m_{H_1^\pm})- F_{Aj}^* F_{Ai} s_\chi^2 F_2(m_{\ell_i}, m_{\ell_j}, m_{H_2^\pm})   \right], \\
(a_{L}^{H^\pm_1})_{ij} &= \frac{1}{16 \pi^2} \sum_{A=1}^4 \left[ (Y_{\Phi'})_{Aj}^{*} (Y_{\Phi'})_{Ai} c_\chi^2 F_2(m_{\ell_i}, m_{\ell_j}, m_{H_1^\pm})- F_{Aj}^* F_{Ai} s_\chi^2 F_1(m_{\ell_i}, m_{\ell_j}, m_{H_2^\pm})   \right], \\
(a_{R}^{H^\pm_2})_{ij} &= (a_{R}^{H^\pm_1})_{ij}\Big|_{c_\chi^2 \leftrightarrow s_\chi^2}, \quad (a_{L}^{H^\pm_2})_{ij} = (a_{L}^{H^\pm_1})_{ij}\Big|_{c_\chi^2 \leftrightarrow s_\chi^2}. 
\end{align}
The loop functions are defined by 
\begin{equation}
F_{1[2]} (m_1, m_2, m_3) = \int [dX] \frac{x z m_2 [x y m_1]}{(x^2-x) m_1^2 + x z (m_1^2 - m_2^2)  + (y+z) m_3^2},
\end{equation}
where we write $\int [dX] \equiv \int_0^1 dx dy dz \delta(1-x-y-z) $. 
The contributions from diagrams with the neutral scalar bosons ($\varphi = h, H, A$) inside loop are calculated as
\begin{align}
& (a_{R}^\varphi)_{ij} = \frac{1}{8 \pi^2} \sum_{A=1}^4 \int [dX] \frac{xy m_{\ell_i} f^{*jA}_\varphi f^{Ai}_\varphi + x z m_{\ell_j} g^{*jA}_\varphi g^{Ai}_\varphi + (1-x) m_{\ell_A} f^{*jA}_\varphi g^{Ai}_\varphi }{-x(1-x) m^2_{\ell_i} - x z (m^2_{\ell_j} - m^2_{\ell_i}) + (z+ y) m^2_{\ell_A} + x m_\varphi^2}, \\
& (a_{L}^\varphi)_{ij} = \frac{1}{8 \pi^2} \sum_{A=1}^4 \int [dX] \frac{xz m_{\ell_i} f^{*jA}_\varphi f^{Ai}_\varphi + x y m_{\ell_j} g^{*jA}_\varphi g^{Ai}_\varphi + (1-x) m_{\ell_A} g^{*jA}_\varphi f^{Ai}_\varphi }{-x(1-x) m^2_{\ell_i} - x z (m^2_{\ell_j} - m^2_{\ell_i}) + (z+ y) m^2_{\ell_A} + x m_\varphi^2}, 
\end{align}
where the couplings corresponding to the neutral scalars $\varphi$ are given by 
\begin{align}
& f^{AB}_h = \frac{1}{\sqrt{2}} (Y_\Phi^*)_{BA} s_{\beta-\alpha} +  \frac{1}{\sqrt{2}} (Y_{\Phi'}^*)_{BA} c_{\beta-\alpha}, 
\quad g^{AB}_h =\frac{1}{\sqrt{2}}  (Y_\Phi)_{AB} s_{\beta-\alpha}  + \frac{1}{\sqrt{2}}   (Y_{\Phi'})_{AB} c_{\beta-\alpha} \nonumber \\
& f^{AB}_H = \frac{1}{\sqrt{2}} (Y_\Phi^*)_{BA} c_{\beta-\alpha} - \frac{1}{\sqrt{2}}  (Y_{\Phi'}^*)_{BA} s_{\beta-\alpha}, 
\quad  g^{AB}_H = \frac{1}{\sqrt{2}}  (Y_\Phi)_{AB} c_{\beta-\alpha}  - \frac{1}{\sqrt{2}}   (Y_{\Phi'})_{AB} s_{\beta-\alpha}  \nonumber \\
& f^{AB}_A = - \frac{i}{\sqrt{2}} (Y_{\Phi'}^*)_{BA} c_{\beta-\alpha}, \quad g^{AB}_A = \frac{i}{\sqrt{2}} (Y_\Phi)_{AB} c_{\beta-\alpha}. 
\end{align}

\section{One-loop Effective Potential at Finite Temperature\label{sec:ap1}}
\label{sec:finiteT}

In order to discuss the one-loop effective potential at finite temperature, 
we introduce the classical constant field configurations $\vec{\varphi^{}}=(\varphi_1,\varphi_2)$ where $(\varphi_1,\varphi_2)|_{T=0}\equiv(v_1,v_2)$ at the VEVs of our present vacuum.
We here do not consider the possibility of CP-breaking and/or charge-breaking vacua for simplicity. 

The effective potential receives additional contributions from thermal loop diagrams, and is modified to~\cite{Dolan:1973qd}
\begin{align}
  V_{\text{eff}}(\vec{\varphi},T)
    =
      V_\text{tree}(\vec{\varphi^{}})
      +V_\text{CW}[M_i(\vec{\varphi^{}},T)]
      +V_T\left[M_i(\vec{\varphi^{}},T)\right], 
\end{align}
with
\begin{align}
  V_\text{tree}(\vec{\varphi^{}})
    &\equiv
      \frac{m_1^2}{2} \varphi_1^2+\frac{m_2^2}{2} \varphi_2^2 -m_3^2 \varphi_1 \varphi_2+\frac{\lambda_1}{8} \varphi_1^4 + \frac{\lambda_2}{8} \varphi_2^4+\frac{1}{4}(\lambda_3+\lambda_4) \varphi_1^2 \varphi_2^2, \\
  V_\text{CW} \left[M_i(\vec{\varphi})\right]
    &\equiv
      \sum_{i} \frac{n_i}{64\pi^2} 
      M_i^4(\vec{\varphi})  \left(\ln \frac{M_i^2(\vec{\varphi})}{Q^2}-c_i\right), \\
  V_T\left[M_i(\vec{\varphi^{}})\right]
    &\equiv
      \frac{T^4}{2\pi^2} 
        \left\{ \sum_{i=\text{bosons}} 
        n_i I_\text{B}\left( \frac{ M^2_i(\vec{\varphi^{}})}{T^2} \right)+ \sum_{i=\text{fermions}} n_i I_\text{F}\left( \frac{ M^2_i(\vec{\varphi^{}})}{T^2} \right) \right\} ,
    \label{eq:effpot_massless}
\end{align}
where $n_i$ and $M_i(\vec{\varphi^{}})$ denote the degrees of the freedom and the field-dependent masses for particles 
$i=h,~G^0,~H,~A,~G^\pm,~H_1^\pm,~H_2^\pm,~\gamma_{L, T}^{},~Z_{L, T}^{},~W_{L, T}^{\pm},~t$ and $b$, respectively. Namely, 
\begin{align}
&  n_{h,G^0,H,A}=1,\,
  n_{G^\pm,H_1^\pm,H_2^\pm}=2, \notag\\
&  n_{\gamma_L,Z_L}=1,\,
  n_{W_L^\pm}=2,\,
  n_{\gamma_T,Z_T}=2,\,
  n_{W_T^\pm}=4,\,
  n_{t,b}=-12.
\end{align}
The renormalization scale $Q$ is set at $v$ in our analysis.
We take the $\overline{\rm MS}$ scheme, where the numerical constants $c_i$ are determined to be $3/2$ ($5/6$) for scalars and fermions (gauge bosons).
The thermal correction is given by $I_{B,F}(a_i^2)= \int^{\infty}_0 dx \ x^2 \ln \left[1 \mp \exp (-\sqrt{x^2+a_i^2}) \right]$ for bosons $(-)$  and fermions $(+)$, respectively.

In order to take ring-diagram contributions into account, we have introduced the field-dependent masses depending on the temperature in the effective potential by \cite{Carrington:1991hz}
\begin{align}
  M_i^2(\vec{\varphi^{}}, T)
  \equiv M_i^2(\vec{\varphi^{}})+\Pi_i(T),
\end{align}
where $\Pi_i^{}(T)$ denote the finite temperature contributions to the self energies of the fields $i$.

The thermally corrected field-dependent masses of the Higgs bosons are
\begin{align}
  M^2_{h,H}(\varphi_1^{},\varphi_2^{},T), \quad
  M^2_{G^0,A}(\varphi_1^{},\varphi_2^{},T), \quad
  M^2_{G^\pm,H_1^\pm,H_2^\pm}(\varphi_1^{},\varphi_2^{},T),
\end{align}
which are obtained by diagonalizing the following mass matrices,
\begin{align}
  M_\text{CP-even}^2 
  &=
  \frac{1}{4}
  \begin{pmatrix}
  4 m_1^2 + 6 \lambda_1 \varphi_1^2 + 2 (\lambda_3+\lambda_4) \varphi_2^2 & - 4 m_3^2 + 4 (\lambda_3+\lambda_4) \varphi_1 \varphi_2 \\
  - 4 m_3^2 + 4 (\lambda_3+\lambda_4) \varphi_1 \varphi_2 & 4 m_2^2 + 6 \lambda_2 \varphi_2^2 + 2 (\lambda_3+\lambda_4) \varphi_1^2 \\
  \end{pmatrix} + \Pi_N(T),  \label{tm1}\\
  M_\text{CP-odd}^2
  &=
  \frac{1}{4}
  \begin{pmatrix}
  4 m_1^2 + 2 \lambda_1 \varphi_1^2 + 2(\lambda_3+\lambda_4) \varphi_2^2 & - 4 m_3^2 \\
  - 4 m_3^2 & 4 m_2^2 + 2 \lambda_2 \varphi_2^2 + 2 (\lambda_3+\lambda_4) \varphi_1^2 \\
  \end{pmatrix} + \Pi_N(T),  \label{tm2}\\
  M_\text{Charged}^2
  &=
  \frac{1}{4}
  \begin{pmatrix}
  4 m_1^2 + 2 \lambda_1 \varphi_1^2 + 2 \lambda_3 \varphi_2^2 & - 4 m_3^2 + 2 \lambda_4 \varphi_1 \varphi_2 & 2\sqrt{2} \mu \varphi_2 \\
  - 4 m_3^2 + 2 \lambda_4 \varphi_1 \varphi_2 & 4 m_2^2 + 2 \lambda_2 \varphi_2^2 + 2 \lambda_3 \varphi_1^2 & -2\sqrt{2} \mu \varphi_1 \\
  2\sqrt{2} \mu \varphi_2  & -2\sqrt{2} \mu \varphi_1 & 4 m_S^2 + 2 \sigma_1 \varphi_1^2 + 2 \sigma_2 \varphi_2^2
  \end{pmatrix} + \Pi_C(T), \label{tm3}
\end{align}
where $\Pi_N(T)$ and $\Pi_C(T)$ are the thermal corrections to the Higgs boson masses at $T$ given as
\begin{align}
  \Pi_N(T)
  &=
  \frac{T^2}{12} \text{diag}(N_1,N_2), \quad
  \Pi_C(T)
  =
  \frac{T^2}{12} \text{diag}(C_1,C_2, C_3), \\
  N_1
  &=C_1=
  3 \lambda_1 + 2 \lambda_3 + \lambda_4 + \sigma_1+ \frac{3}{4} (3 g^2 + g'^2), \\
  N_2
  &=C_2=
  3 \lambda_2 + 2 \lambda_3 + \lambda_4 + \sigma_2+ \frac{3}{4} (3 g^2 + g'^2) + 3 (y_t^2+y_b^2), \\
  C_3
  &=
  2 \sigma_1 + 2 \sigma_2.
\end{align}
Here, $g$ and $g'$ ($y_t^{}$ and $y_b^{}$) are the gauge couplings of $SU(2)_L$ and $U(1)_Y^{}$ (the top and bottom Yukawa couplings), respectively.
We note that the $m_1^2$ and $m_2^2$ parameters appearing in Eqs.~(\ref{tm1}), (\ref{tm2}) and (\ref{tm3}) are determined by solving the tadpole conditions at the 
zero temperature. 

\vspace*{4mm}

\bibliography{references}

\end{document}